\documentclass[sn-jnl]{sn-jnl}

\usepackage{amsmath,amssymb}
\usepackage{float}
\usepackage[table,dvipsnames]{xcolor}

\usepackage{multicol,multirow}
\usepackage[
  style=apa,
  backend=biber,
  sortcites=true,
  sorting=nyt
]{biblatex}
\DeclareLanguageMapping{english}{english-apa}

\usepackage{comment}
\usepackage{array}
\usepackage{booktabs}
\usepackage{longtable}
\usepackage{changepage}
\usepackage{adjustbox}
\usepackage{tikz}
\usetikzlibrary{positioning, arrows.meta}

\definecolor{hrraicyan}{RGB}{197, 220, 254} 
\definecolor{hrraigreen}{HTML}{008a5e}

\newcommand{\hrrai}[1]{{\texttt{HR-RAI}}}

\usepackage{verbatim}

\usepackage{soul}

\begin{document}

\title[The HR-RAI Toolkit]{On the Challenges of Using LLMs in High-Risk AI Recruitment: Guiding Questions for Legal Conformity with the AI Act }

\author[1,3]{\fnm{Clara} \sur{Punzi}}
\equalcont{These authors contributed equally.}

\author[2,3]{\fnm{Stefano} \sur{Tramacere}}
\equalcont{These authors contributed equally.}

\author[3]{\fnm{Andrea} \sur{Fedele}}
\equalcont{These authors contributed equally.}

\author[1]{\fnm{Sezer} \sur{Kutluk}}
\author[1]{\fnm{Roberto} \sur{Pellungrini}}
\author[1]{\fnm{Gizem} \sur{Gezici}}
\author[4]{\fnm{Paolo} \sur{Baracco}}
\author[4]{\fnm{Laura} \sur{Castello}}
\author[4]{\fnm{Eleonora} \sur{Cusin}}
\author[4]{\fnm{Eduardo} \sur{Costa}}
\author[4]{\fnm{Amedeo} \sur{Giacovelli}}
\author[4]{\fnm{Mario} \sur{D'Almo}}
\author[1]{\fnm{Fosca} \sur{Giannotti}}
\author[3]{\fnm{Dino} \sur{Pedreschi}}
\author*[1]{\fnm{Marzio} \sur{Di Vece}}\email{marzio.divece@sns.it}

\affil[1]{\orgname{Scuola Normale Superiore}, \orgaddress{\city{Pisa}, \country{Italy}}}
\affil[2]{\orgname{Scuola Superiore Sant'Anna}, \orgaddress{\city{Pisa}, \country{Italy}}}
\affil[3]{\orgname{Università di Pisa}, \orgaddress{\city{Pisa}, \country{Italy}}}
\affil[4]{\orgname{Intesa Sanpaolo}, \orgaddress{\city{Turin}, \country{Italy}}}

\abstract{

The EU Artificial Intelligence Act (AI Act) imposes detailed obligations on providers of high-risk AI systems but offers limited operational guidance, particularly when these systems integrate third-party General-Purpose AI (GPAI) models. This paper introduces a toolkit that fosters the development of legal-by-design AI-assisted recruitment systems by translating the core legal obligations of the AI Act's conformity assessment---specifically Articles 9--15 and Annex IV---into techno-normative requirements and corresponding guiding questions. While recruitment pipelines may involve components with varying technical complexity and differing levels of algorithmic transparency, this paper focuses on the most challenging case: the integration of third-party GPAI models into critical stages of the selection process, including resume and job description parsing, applicant scoring, and shortlist generation. This scenario presents significant challenges for downstream providers of high-risk AI systems, as it underscores the difficulties in obtaining the technical information needed to demonstrate compliance with the AI Act's techno-normative requirements, both prior to and during market deployment. After introducing the toolkit and illustrating its application through a simplified example, the paper explores informational dependencies between upstream and downstream GPAI model providers. These dependencies may produce informational gaps across the AI supply chain, potentially hindering the collection of evidence necessary for the AI system provider to ensure legal compliance. The resulting lack of technical transparency negatively affects the deployer, who may struggle to maintain meaningful oversight of the system, and ultimately impacts the persons subject to the AI output.
}


\maketitle
\section{Introduction} 
\label{sec:intro}


Preventing the harmful effects of Artificial Intelligence (AI) systems has become a major concern,
especially in high risk domains such as recruitment, education, credit scoring and insurance, where automated decision systems directly affect persons’ rights and access to opportunities \parencite{Stazi2025,Deshpande24}. The European Union AI Act \parencite{AIAct} addresses this issue by imposing detailed obligations on providers and deployers of high-risk AI systems~~\parencite{AIAct}. Yet, a persistent challenge remains: how to practically implement these abstract legal provisions during AI systems' design, deployment, monitoring, and assessment \parencite{Duke2025}.

This paper addresses this challenge through collaborative work between academic researchers and key stakeholders from
 a partner banking institution, namely its AI software developer and compliance teams.  It proposes a toolkit, namely \textbf{H}igh \textbf{R}isk - \textbf{R}esponsible \textbf{A}rtificial \textbf{I}ntelligence (\hrrai{}), intended to facilitate the development of an abstract legal-by-design AI-assisted recruitment system. This system integrates third-party General-Purpose AI (GPAI) models at multiple critical stages of the pipeline, without the need for additional training or fine-tuning.
Applications used in the recruitment context are included among the high-risk sectors listed in Annex III(4)(a), as explicitly referenced by Article 6(2). Under Article 8, providers of AI systems are required to ensure compliance with the techno-normative requirements set forth in Section 2, with the aim of ensuring, in accordance with the precautionary principle~\parencite{Garnett2017}, an adequate level of product safety to protect the health and fundamental rights of persons. 
The pipeline examined in this work presents heightened compliance challenges due to its reliance on 
GPAI components characterized by multiple levels of opacity. These range from corporate or institutional secrecy in the case of closed models to algorithmic inaccessibility, even for open-source models \parencite{Buttaboni2026}. As such, the scenario serves as a stress test for the difficulties that providers may encounter when gathering evidence to demonstrate compliance, with a particular focus whenever closed components are integrated.
The choice underlying this study reflects prevailing industry practices shaped on one hand by market and financial pressures to adopt the latest available technologies, on the other by benchmark studies showing that proprietary models continue to outperform open-source alternatives in terms of performance, user experience, access to state-of-the-art capabilities \parencite{epoch2024openmodelsreport}, and perceived reliability \parencite{huang2024trustllm}.

The proposed toolkit supports the evaluation of high-risk recruitment AI systems against the requirements of the AI Act, translating the legal obligations set out in Chapter III, Section 2 of the Act into a structured checklist of techno-normative requirements and corresponding guiding questions. These questions are intended to stimulate critical reflection on the adequacy of an AI system’s design, development and deployment. 
In doing so, the toolkit serves a dual function: first, to assist in the systematic collection of evidence necessary for compliance verification; and second, to encourage responsible practices among AI designers and developers \parencite{Bignami2025}. In this way, the toolkit not only facilitates compliance assessment but also promotes the integration of ethical and legal considerations throughout the AI development lifecycle, thereby advancing the implementation of transparent and legal-by-design AI systems. 

\begin{figure}[t!]
    \centering
    \includegraphics[width=\textwidth]{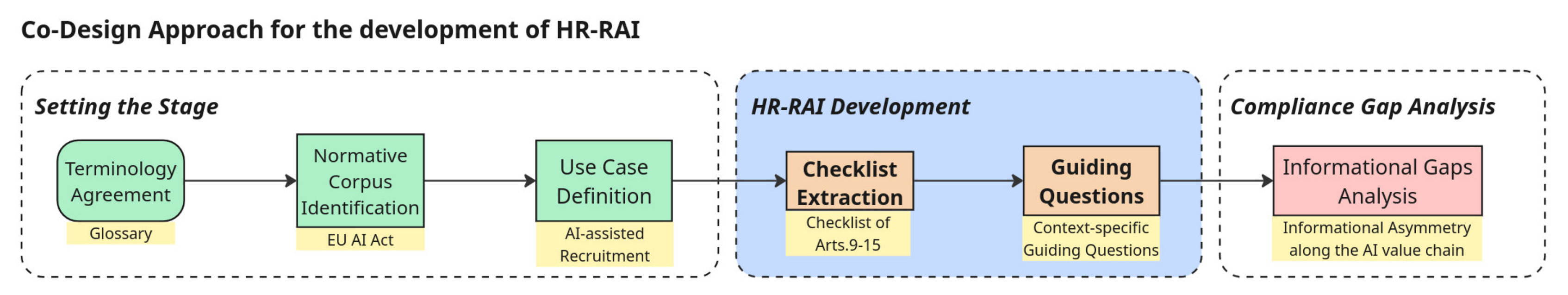}
    \caption{\textbf{Overview of the co-design methodology.} 
     The three phases of the methodological framework: (1) \textit{Setting the stage}, establishing shared terminology and anchoring the analysis to a concrete deployment scenario: an AI-assisted recruitment system integrating a third-party GPAI model across parsing, representation learning, and scoring stages; (2) \textit{HR-RAI development}, translating legal provisions into guiding questions through iterative exchanges between researchers and stakeholders; and (3) \textit{Compliance gap analysis}, identifying residual informational deficits generated by GPAI integration, examined at the intersection of the AI Act and non-binding instruments such as the EU GPAI Code of Practice \parencite{CodeOfPractice25}.}
    \label{fig:maps_co-design}
\end{figure}

The development of \hrrai{} followed a co-design approach involving academic researchers and compliance and development teams within a partner banking institution. The overall methodological framework, illustrated in Figure~\ref{fig:maps_co-design}, unfolds in three phases: (1) \textit{setting the stage}, establishing shared terminology and anchoring the analysis to a concrete scenario in which an AI-assisted recruitment system integrates a third-party GPAI model; (2) \textit{toolkit development}, translating legal provisions into guiding questions; and (3) \textit{compliance gap analysis}, examining residual challenges stemming from proprietary GPAI integration by analyzing the AI Act alongside non-binding instruments.

The remainder of the paper is structured as follows. 
Section~\ref{sec:legalBackground} sets out the legal framework governing high-risk AI systems under the AI Act, focusing in particular on Articles~9--15 and Annex~IV, and reviews prior efforts in putting legal requirements into practice, including existing tools aimed to facilitate legal-by-design development. 
Section~\ref{sec:relatedTechnical} introduces the application domain by outlining AI-assisted screening techniques in recruitment and the key stages of state-of-the-art technical screening pipelines, situating the regulatory analysis within a concrete deployment context. 
Section~\ref{sec:GuidingQuestions} presents the core contribution of the paper, namely the \hrrai{} toolkit. It details the checklist of techno-normative requirements and guiding questions that translate Articles~9--15 and Annex~IV into actionable legal-by-design questions, and, in Section \ref{sec:examples}, 
presents a simplified example of how the toolkit can be used in practice by providers. 
In Section \ref{sec:challenges}, we also discuss informational gaps along the AI value chain, highlighting how operational compliance is shaped by structural asymmetries when downstream providers integrate third-party GPAI models. 
Finally, Section~\ref{sec:conclusions} concludes by summarising the contribution, outlining the main limitations, and indicating directions for future work, including empirical and quantitative evaluation.

\section{Legal Background: Techno-Normative Requirements in the AI Act} \label{sec:legalBackground} 

Early governance efforts, such as principles for trustworthy or responsible AI~\parencite{EthicsGuidelines2019,OECD,PAI,AIBluePrint,Dafoe2018} placed emphasis on voluntary commitments aimed at preventing AI-related risks. While influential, these soft-law instruments exhibit limited practical effectiveness and, in some cases, may even prove problematic. Their non-binding nature grants broad discretion to private and public entities, thereby carrying the risk of fragmented and uneven implementation, as well as the  potential for harm to persons subjected to such systems, due to violation of their fundamental rights~\parencite{comande2020unfolding}.  In response, the European Union (EU) has opted for a fully binding hard-law approach by adopting a horizontal regulation on AI, known as the AI Act~\parencite{AIAct}. 
Its primary legal basis is Article 114 of Treaty on the Functioning
of the EU (TFEU), which pertains to the functioning of the internal market, supplemented by Article 16 TFEU concerning the protection of personal data. This regulatory choice is clearly articulated in Article 1 of the Regulation, which establishes two main objectives: first, to improve the functioning of the EU market while promoting the adoption of human-centric and trustworthy AI; and second, to ensure a high level of protection for health, safety, and fundamental rights, as enshrined in the Charter of Fundamental Rights, including democracy, the rule of law, and environmental protection, against the harmful effects of AI systems.

The AI Act adopts a risk-based approach to AI regulation \parencite{DeGregorio2022}. It distinguishes between prohibited AI practices posing unacceptable risks (Chapter II), high-risk AI systems subject to the majority of regulatory requirements (Chapter III), and limited-risk systems subject to specific transparency obligations (Chapter IV). In addition, the Act introduces specific provisions governing general-purpose AI models, which can be deployed across a wide range of tasks (Chapter V). However, despite the adoption of an \textit{ex ante}, risk-based regulatory framework, certain features of traditional product safety regimes remain difficult to reconcile with the objective of ensuring that AI systems do not effectively violate fundamental rights~\parencite{almada2025euaiact}. This tension becomes particularly evident in the context of high-risk AI systems, as defined in Article 6 and Annexes I and III, which may be deployed in both public and private sectors~\parencite{fra2025}. To mitigate these risks and ensure more effective protection of fundamental rights (Recital 48), the AI Act adopts a\textit{ ‘by design’} approach for high-risk AI systems, requiring the integration of techno-normative safeguards during the design and development of AI systems~\parencite{Cleynenbreugel_2021}. This approach obliges providers to ensure compliance with legal norms from the design stage~\parencite{hildebrandt_boundary_2023}.

High-risk systems include AI applications intended to be used, among others, for biometric identification (Annex III, 1(a)), educational training (Annex III, 2(a)), migration management (Annex III, 7), and, as examined in this study, the recruitment or selection of persons in the employment sector 
(Annex III, 4(a))\footnote{Annex III, 4(a) states that \textit{``High-risk AI systems intended to be used for the recruitment or selection of natural persons, in particular to place targeted job advertisements, to analyse and filter job applications, and to evaluate candidates''}}. The techno-normative requirements primarily applicable to providers of high-risk AI systems are laid down in Articles 9 to 15 of Chapter III, Section 2, forming the core of the AI Act’s conformity assessment (Article 43). The Act codifies a comprehensive set of technical and organizational safeguards aimed at ensuring an adequate level of legal protection. These include: first, the implementation of a risk management system (Article 9); second, the adoption of appropriate data quality and governance measures (Article 10); third, the preparation of detailed technical documentation in line with the Annex IV (Article 11); fourth, the establishment and maintenance of an ongoing record-keeping system (Article 12); fifth, the fulfillment of transparency and information provision towards deployers, including the interpretability of outputs (Article 13); sixth, the incorporation of human oversight mechanisms to mitigate risks that may amplify deployer biases and ensure awareness of the decisions made (Article 14); and, finally, the implementation of measures designed to ensure accuracy, robustness, and cybersecurity (Article 15).

Given that the text has a horizontal scope of application, its provisions must be implemented with due regard to the specificities of different sectoral contexts, in coordination with other relevant legal and organisational frameworks\footnote{For instance, where an AI system is deployed within recruitment and employment processes, the AI Act must operate in conjunction with the applicable labor, employment, and anti-discrimination law.}. In particular, developers must be aware that, from an operational standpoint, compliance with the technical requirements established by the Regulation constitutes only a first level of conformity. Full compliance, in fact, also depends on the adoption of organisational procedures and governance arrangements tailored to the specific characteristics of the sector in which the system is deployed.

In light of the foregoing, the transparency principle plays a central role in facilitating this initial level of conformity and in strengthening accountability mechanisms among the various actors involved~\parencite{Kaminski_2020}. It constitutes a \textit{fil rouge}, finding expression across the multiple techno-normative requirements applicable to the development and deployment of high-risk AI systems, albeit in different forms~\parencite{Tramacere25}. Taken together, these legal and technical instruments contribute to a clearer allocation of responsibilities along the AI value chain and enable deployers and competent authorities to exercise more effective oversight. In doing so, their operationalization can contribute to strengthening the effective implementation of the right to a meaningful explanation by the deployer for persons subject to high-risk AI systems, as provided for in Article 86.

The importance of transparent design and development is also evident in high-risk AI systems 
 that integrate GPAI models, including third-party LLMs, as upstream components~\parencite{Liao2024AI}. In this regard, the AI Act establishes specific transparency obligations for GPAI model providers, as outlined in Articles 53. It is important to emphasise that when GPAI providers also act as providers of high-risk AI systems, they are required to comply with a dual set of obligations: those related to the development of GPAI models (Chapter V) and those applicable to the development of high-risk systems (Chapter III). When, on the other hand, a GPAI model is made available to third parties who integrate it into their own architecture to develop a high-risk AI system, a distinction arises between upstream and downstream providers\footnote{Article 3(68) states that \textit{`` `downstream provider' means a provider of an AI system, including a general-purpose AI system, which integrates an AI model, regardless of whether the AI model is provided by themselves and vertically integrated or provided by another entity based on contractual relations''}.}. In this case, the responsibility for ensuring final compliance with the requirements for high-risk systems lies with the downstream provider of the high-risk AI system~\parencite{hacker2025finetuning}. Therefore, in order for compliance to be effectively ensured along the entire value chain, it is essential, as discussed in section~\ref{sec:challenges}, that the downstream provider has access to the adequate technical information provided by the upstream provider, to prevent informational gaps from compromising the overall reliability of the system.

\paragraph{Responsible AI 
 practices in recruitment} \label{sec:relatedComplianceTools}

The AI Act establishes binding obligations for high-risk AI system providers, but its provisions inevitably leave room for interpretation in practical implementation. Compliance tools, such as guidelines and checklists, play a crucial role in translating the techno-normative requirements into operational, reproducible, and verifiable procedures. These tools can facilitate technical implementation, enhance accountability, reduce interpretive uncertainty, and contribute to the standardisation of certain
choices throughout the whole AI lifecycle,
thereby bridging the disciplinary gaps between legal and technical experts.
In this regard, the literature highlights a growing body of studies exploring how to translate the techno-normative requirements of the AI Act into technical and organisational safeguards, adapting them to various high-risk application contexts \parencite{Madaio20,Constantinides24,Bignami2025}. Among these, recruitment stands out as a prominent area of broad interest, also serving as the applicative focus of this study.

The UK Department for Science, Innovation \& Technology’s Responsible AI in Recruitment guidance \parencite{RaiUK24} identifies key ethical risks associated with AI-enabled recruitment and outlines AI assurance mechanisms, such as evaluation metrics, risk management processes, and governance practices, to support organisations in procuring and deploying recruitment AI systems in compliance with applicable legal requirements.
Moreover, within the EU project \textit{``FINDHR – Fairness and Intersectional Non-Discrimination in Human Recommendation''}, an interdisciplinary consortium from academia, industry, and civil society released various resources to
counteract algorithmic discrimination in AI recruitment systems\footnote{Available at \url{https://findhr.eu/toolkits/}}. These resources encompass specific guidelines and recommendations for software developers, human resources and policymakers, as well as open domain-specific datasets, tools and training programmes for professionals.

\section{AI-Assisted Screening in Recruitment} 
\label{sec:ai_hiring}

\begin{figure}[t!]
    \centering
    \includegraphics[width=\linewidth]{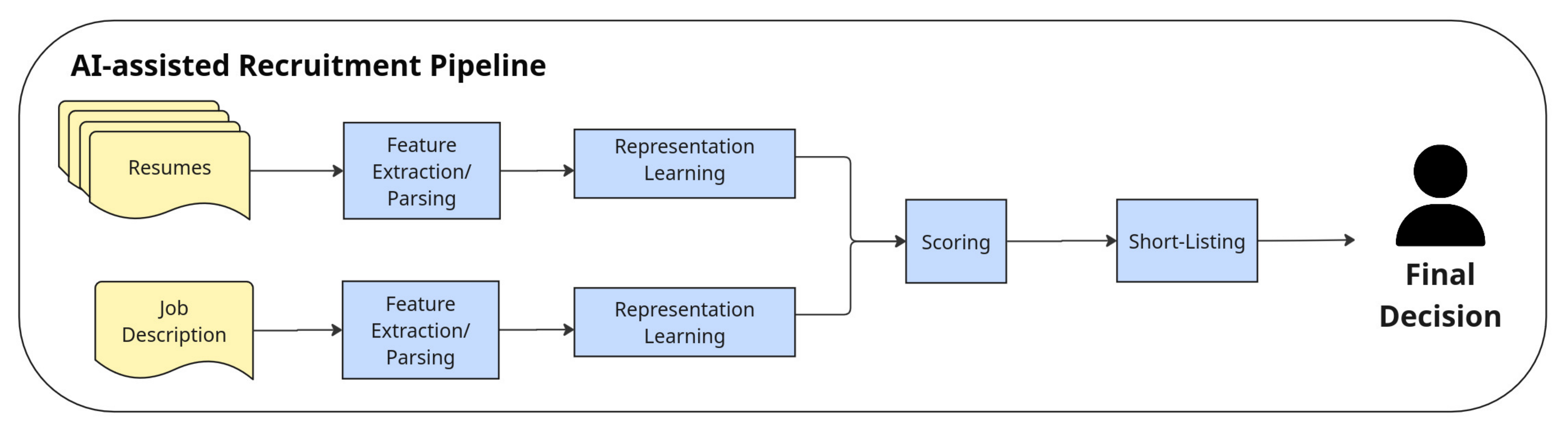}
    \caption{\textbf{AI-assisted recruitment pipeline.} The pipeline comprises five stages: (1) \textit{parsing} extracts structured information from unstructured CVs and job descriptions; (2) \textit{representation learning} encodes this information into vector representations; (3) \textit{scoring} computes similarity metrics between candidate profiles and job requirements; (4) \textit{short-listing} ranks candidates based on scores; and (5) \textit{final decision} remains with HR professionals.}
    \label{fig:recruitment_pipeline}
\end{figure}

The recruitment process encompasses all activities following the collection of candidate applications, including screening and interviewing procedures leading to the final selection.
Over the last decades, a variety of Applicant Tracking Systems (ATS) have been developed to support Human Resources (HR) departments, and, in recent years, these systems have increasingly incorporated AI-based components~\parencite{cheng2021critical,boyd2025interdisciplinary}.
These AI systems are designed to assist HR professionals in the screening phase by ranking applicants based on the comparison between their resumes and the job descriptions provided by the company. 
In this way, the labor intensive task of screening large volumes of resumes is alleviated, while the final recruitment decision remains the responsibility of the human recruiter~\parencite{vaishampayan2025human}.
Despite the lack of general agreement in the literature regarding a standardised
nomenclature for the tasks involved in AI-assisted screening, in this work we focus on the following stages shown in Fig.~\ref{fig:recruitment_pipeline}: \textit{parsing}, \textit{representation learning}, \textit{scoring}, \textit{short-listing}, and \textit{final decision}.

\textit{Parsing} refers to the extraction of structured information from unstructured resume and job description documents. This task remains challenging due to the free-form nature of human-authored documents, which often exhibit heterogeneous layouts, phrasing, and visual elements. 
The extracted text is then encoded into vector representations through 
\textit{representation learning}, enabling the matching of applicants' attributes against job requirements.
\textit{Scoring} of applicants is performed using similarity metrics, such as cosine similarity, or directly through an LLM recommendation, and applicants are ranked according to their estimated suitability for the position to produce a \textit{shortlist} to be reviewed by recruiters.
\textit{Final recruitment decisions} remain with HR professionals to ensure accountability and regulatory compliance, as AI-assisted screening for recruitment is classified as a high-risk application under the AI Act. 

\paragraph{Related Work}\label{sec:relatedTechnical}
Only a limited number of studies explicitly treat \emph{parsing} as a standalone component of automated screening pipelines; for example, \textcite{bhatia2019end} extract structured information from unstructured PDF documents and represent it in a tabular format.
Instead, the dominant emphasis in the literature is on \textit{representation learning}, namely encoding resumes and job descriptions into comparable representations that support downstream matching and scoring \parencite{zhang2010using, ajjam2025ai}.
Research in this area can be grouped into three paradigms: sparse lexical representations, dense vector representations, and LLM-based end-to-end approaches. These paradigms also differ markedly in the \emph{auditability} of the resulting pipeline.

A first line of work relies on sparse representations such as TF--IDF, where screening is largely implemented through weighted keyword overlap \parencite{mishra2015analysis, zhang2010using}. While comparatively transparent, these approaches are limited in capturing semantic meaning.
Most contributions instead adopt dense embeddings, mapping text into continuous vector spaces in which similarity can be computed and optimised \parencite{elsafty2018document, zhao2021embedding, dave2018combined}. Early dense pipelines are built on static embeddings, including Word2Vec \parencite{mikolov2013distributed}.
Subsequent work has increasingly leveraged pretrained language models (PLMs), notably BERT \parencite{devlin2019bert}, to derive contextualised representations.
Representative approaches include BERT-based encoders \parencite{bhatia2019end, decorte2021JobBERT, gnehm2022evaluation}, sentence-level similarity models based on SBERT \parencite{reimers2019sentence} as adopted in \parencite{lavi2021consultantbert, rosenberger2025careerbert}, and multilingual encoders such as XLM-RoBERTa \parencite{conneau2020unsupervised} for cross-lingual job-market applications \parencite{zhang2023escoxlm}.

Only a small number of studies explicitly incorporate LLMs \parencite{li2023skillgpt, vaishampayan2025human}. These works move toward end-to-end formulations in which a single model interaction can subsume multiple stages, including information extraction, representation, and scoring, through prompting rather than task-specific training. Specifically, \textcite{vaishampayan2025human} utilize a closed LLM for an end-to-end pipeline for candidate rating through prompt engineering techniques. In contrast, although it does not propose a full recruitment system, \textcite{li2023skillgpt} utilize an open LLM backbone in a skill extraction and standardisation pipeline.
While appealing, this shift substantially complicates conformity assessment. 
Regardless of whether the LLM is open- or closed-source, its use across one or more stages of the pipeline raises important concerns.
Generally, by relying on opaque models, the access to transparency-relevant information needed to interpret the system’s internal processes and outputs is restricted. 
Furthermore currently most widely adopted models are kept closed by their developing company, leading to private code and data that cannot be accessed by external parties.

\section{\hrrai{}: A compliance toolkit for AI-assisted recruitment}
\label{sec:GuidingQuestions} 


In this section we describe \hrrai{}, a compliance toolkit designed to 
guide providers of AI-assisted recruitment tools towards legal-by-design solutions.
The toolkit provides a reasoned list of questions designed to facilitate the implementation of the legally binding provisions that constitute the technical backbone of the high-risk regime under the AI Act, namely Articles~9–15 and Annex~IV. The provisions establishes a set of techno-normative requirements that, in part, codify existing best practices in responsible AI development, while also forming the core of the conformity assessment procedure laid down in Article 43. The toolkit provides a structured analysis through two interconnected levels:
\begin{itemize}
    \item The first level consists of a checklist obtained by decomposing each article into granular sub-requirements, as consolidated in the literature regarding compliance practices that translate legal duties into actionable control points \parencite{Altai2020,Constantinides24,Madaio20,Bignami2025}. 
    \item The second level breaks down each item of the checklist into one or more guiding questions focused on technical implementation details and organisational procedures. These questions were conceived to serve a dual purpose: (i) to assist in the systematic collection of evidence necessary for compliance assessment, and (ii) to trigger awareness and encourage reflective practice among AI designers and developers~\parencite{Bignami2025}. 
\end{itemize}
For each of the analysed articles, the toolkit also includes selected excerpts from the corresponding Recitals of the AI Act, providing interpretative guidance to support providers in understanding the legislative rationale and regulatory expectations from the EU perspective.

Overall, the \hrrai{} toolkit covers seven articles (Articles~9–15) and one annex (Annex~IV), resulting in a total of 94 checklist items (approximately 13 per article) and 488 guiding questions (on average nearly 70 per article). The number of guiding questions associated with each requirement varies across articles, ranging from 1 to 22, with an average of five.
 
The \hrrai{} toolkit is intended to be used by both compliance and development teams in a collaborative manner to better achieve legal-by-design AI systems. It provides a structured framework to assess and document the conformity of AI recruitment systems, emphasizing that achieving and demonstrating legal compliance requires technical and organizational measures spanning the entire life-cycle of the AI system. An excerpt from the compliance checklist and guiding questions stemming from Article 10, 2(a) and 2(b) is presented in \autoref{tab:excerptGQ}. 
The full toolkit is freely accessible both in textual form in \autoref{app:hrrai} and through an online web interface\footnote{\hrrai{} toolkit available at: \url{https://anonymous.4open.science/w/HR-AI-compliance-7279}.}.

\begin{table}[!h]
    \centering\footnotesize
    \begin{tabular}{p{4cm}p{8cm}}
\rowcolor[HTML]{cfe2f3}\textbf{Compliance checklist} & \textbf{Guiding questions} \\
& \\
\multicolumn{2}{l}{
\cellcolor[HTML]{efefef}
\parbox{12cm}{
\textit{
2. Data governance and management practices should be appropriate for the intended purpose of the high-risk AI system and concern:
}}
} \\
\multirow{10}{*}{\parbox{4cm}{2(a). \textbf{Design choices for} the \textit{testing data}.}} & 10.2.a.1 Which format did you use for CVs and job descriptions (e.g., structured templates vs. unstructured formats like PDFs)? \\
&10.2.a.2 What assumptions about data quality or availability have been made? \\
&10.2.a.3 Does the data pre-processing pipeline include parsing of unstructured input data? If so, how was it conducted? \\
&10.2.a.4   Which languages are supported for CVs and job descriptions? If applicable, how does translation occur, and how did you verify its quality? \\
&10.2.a.5 In the case of unstructured input data, is the output of the parsing pipeline structured according to a homogeneous template? What information is extracted and why? \\
&10.2.a.6  Why did you choose to use a public and/or private test dataset? \\
&10.2.a.7  Is the test data you chose representative of the real context of use of the system? \\
&10.2.a.8 How did you choose the target labels for testing? \\
&10.2.a.9 How did you choose the evaluation metrics for testing? \\
&10.2.a.10 How did you choose to annotate your testing data, if needed? If humans were involved: \\
&-     How was the selection of human annotators conducted (e.g., in terms of AI literacy and domain expertise)? \\
&-     Do human annotators have an internal or external function within the organization? \\
&-     Did you design the annotation process to ensure the inclusion of a variety of perspectives? \\
\midrule\multirow{7}{*}{\parbox{4cm}{2(b). \textbf{Sources and collection processes for} the \textit{testing data}, including details on the original purpose, in case personal data is involved.}} &10.2.b.1 Are the sources of the testing data comprehensively documented in either internal or external documents? \\
&10.2.b.2   How does the collection process handle data covered by copyright? Is the final collected data protected by copyright? \\
&10.2.b.3  How does the collection process handle potentially illegal (e.g., offensive or non-consensual) data? Does the final collected data contain potentially illegal data? \\
&10.2.b.4  Are the data sources for testing data all public? If not, which private data sources did you collect data from? \\
&10.2.b.5  Do the data sources for testing data contain personal information? If so, did you record their original purpose? Moreover, if special categories of personal data are included, did you comply with Article 10(5)? \\
&10.2.b.6  How is test data tracked or updated? In the case of an update, please provide a detailed account of the sources of additional test data (e.g., feedback loops) and the collection processes. \\
&10.2.b.7  Did you record the data collection procedure, including the computational methods employed for extraction (e.g., discriminative or generative methods)? \\
\bottomrule
\end{tabular}
\caption{\footnotesize \textbf{An excerpt from the toolkit given in \autoref{app:hrrai}, showing example compliance checklist items of Article 10 (2(a) and 2(b)), along with their guiding questions.} The first column shows the checklist items that correspond to each paragraph in the Article (or Annex). In the second column, each guiding question (GQ) is identified using the numbering adopted in the toolkit (see \autoref{app:hrrai}): the first number refers to the corresponding Article (or Annex) of the AI Act; the subsequent numbers and letters indicate the relevant paragraph and item; and the final number is progressive.}
\label{tab:excerptGQ}
\end{table}

\subsection{\hrrai{}: How to use} 
\label{sec:examples}



The primary value of the \hrrai{} toolkit lies in supporting developers and compliance teams in adopting a legal-by-design approach, embedding regulatory requirements into the system architecture from the earliest stages of development.

Since the toolkit has already been filtered to include only the legal obligations relevant to the specific application scenario, the objective of the compliance assessment is to demonstrate that \textit{all} items in the checklist are satisfied.
To this end, developers are asked to provide the teams responsible for the assessment with technical and organisational evidence, such as evaluation results, system documentation or risk analyses, in response to the toolkit’s guiding questions. These questions are intended to be used as inputs during design reviews, documentation activities, or internal compliance audits. By structuring the collection of such evidence, the toolkit facilitates the interaction between compliance and development teams to ensure a legal-by-design solution. 
Importantly, the checklist of requirements and related guiding questions span the entire lifecycle of a high-risk AI system, from design and development to deployment and post-market monitoring. 

\begin{table}[ht!]
    \centering\footnotesize
    \begin{tabular}{p{4cm}p{4cm}p{4cm}}
    \rowcolor{hrraicyan}   
    \multicolumn{1}{c}{(Examples of) Guiding questions} & 
    \multicolumn{1}{c}{Relevance} &
    \multicolumn{1}{c}{(Examples of) Evidence}\\
    \textbf{GQ 9.2.a.1} What known risks to health, safety, or fundamental rights may arise when the AI system is used as intended? & 
    Scoring may introduce risks related to discrimination, unfair exclusion of candidates and opacity in decision-making, 
    ultimately affecting fundamental rights such as equal treatment and non-discrimination in the recruitment process. &
    - Structured assessment of statistically different outputs across demographic or linguistic groups \newline 
    - Prompt-based bias evaluations simulating varied applicant profiles\newline
    - [Open LLMs] Evaluation results on benchmark datasets.
    \\ \midrule
    \textbf{GQ 10.2.d.2}  What explicit assumptions are you making about which candidate features (e.g., education, work history, test results) are predictive of job performance?
    & Such features may implicitly encode societal biases or unjustified assumptions,
which may lead to discriminatory outcomes or indirectly disadvantage certain groups.
    & 
    - Feature selection criteria \newline
    - Prompt templates \newline
    - Reports of ablation tests 
    \newline
    - Evidence by domain experts\newline
    - [Open LLMs] Feature importance analysis.
    \\ \midrule
    \textbf{GQ Annex IV.2.b.5} What is the target objective of the system, how is it computed, and why was it chosen? Do you have visibility on the whole computation process, including the sub-tasks performed by third-party components?
    & Scoring is the core component, yielding the ranking position of the candidate. Transparency about the underlying rationale is crucial to ensure, for instance, the right to receive an explanation of the output provided to applicants and to allow them to challenge it.
    & - Model cards from the third-party LLM provider \newline
    - Justification for selected evaluation metrics (e.g., ranking accuracy, fairness metrics, calibration across groups) \newline
    - [Open LLMs] Scoring algorithms 
    \\ \midrule
    \textbf{GQ 12.2.a.1} What types of system outputs or patterns do you treat as indicators that the system might be presenting a risk [...]?
    & Monitoring of outputs can help detect performance degradation, bias, or other harmful patterns after deployment (e.g., systematic rejection of candidates from certain linguistic backgrounds or unexpected shifts in ranking distributions).
    & - Logging mechanisms capturing inputs, outputs and metadata \newline
    - Bias detection tools applied to system outputs \newline    
    - Automated alerts triggered when metrics (e.g., demographic parity
    ) exceed safety thresholds \newline - [Open LLMs] Periodic test vs. performance benchmarks.
    \\ \midrule
    \textbf{GQ 13.3.b.18} Do the instructions cover the characteristics, capabilities, and limitations of your input data, validation, and testing data, based on the system’s intended purpose?
    & The characteristics of CVs (e.g., data format, language, template, demographic representation) may affect system performance, increasing the risk of biased or unreliable candidate scoring when used in inappropriate contexts.
    & - Description of expected input data and known limitations \newline 
    - Linguistic and demographic coverage of the evaluation data \newline
    - Model cards from the third-party LLM provider \newline
    - [Open LLMs] Description of preprocessing.
    \\ \midrule
    \textbf{GQ 14.4.c.1} What information and interpretability tools (e.g., feature importance, counterfactuals, examples and counterexamples) are provided to enable human overseers to interpret system outputs properly?
    & Recruiters should understand why a candidate received a certain matching score to critically assess, detect errors or biased reasoning produced by the LLM and potentially override the system’s recommendation.  
    & - Output of interpretable techniques applied to the output of the system \newline 
    - [Open LLMs] Output of explainable AI techniques, such as feature attribution methods applied to embeddings.
    \\ \midrule
    \textbf{GQ 15.4.7} Did you identify any event that could trigger errors, faults, or inconsistencies due to errors by third-party systems on which your system depends on? 
    & Failures or changes in a third-party model (e.g., model updates, API outages, changes in output behaviour) may introduce inconsistencies, potentially leading to unfair or erroneous scores.
    & - Version control procedures \newline
    - Fallback mechanisms \newline
    - Monitoring systems detecting anomalous output patterns \newline
    - [Open LLMs] Reports of internal testing pipelines that detect regressions after model changes.
    \\
    \end{tabular}
    \caption{\footnotesize A small subset of guiding questions from the full toolkit is reported (one for each of the Articles 9--15), w.r.t. the scoring functionality of an AI-assisted recruitment system using a third-party LLM. 
    In the first column, guiding questions are given using the toolkit adopted numbering. The second column explains why each question is relevant for AI Act compliance.
    The third column provides examples of evidence that developers may document during system design and development to substantiate compliance.}
    \label{tab:exampleGQ}
\end{table}

For clarity of exposition, \autoref{tab:exampleGQ} presents a simplified example of how the toolkit can be used in practice by providers.
The table focuses on a single functionality of the system, namely, the computation of candidate matching scores, and reports one representative guiding question for each of the covered articles of the AI Act, thereby illustrating the need for a synergistic reading of all these requirements. For each question, we briefly explain its relevance for compliance and provide examples of evidence that developers could document to substantiate their design choices. In the full toolkit (reported in \autoref{app:hrrai}), these guiding questions are referenced through a hierarchical numbering scheme 
\texttt{<<Article.paragraph.subparagraph.question number>>};
for instance, the wording \textbf{GQ 13.3.b.18} refers to the \textit{eighteenth} guiding question (\textit{GQ}) related to \textit{Article 13, paragraph (3), subparagraph (b)}.

The evidence column in the Table, illustrates the type of technical measures, documentation or evaluation report that may be collected during system development and use; where relevant, examples specific to open LLMs are marked separately, reflecting the different levels of transparency available when integrating open versus closed models.
For instance, when addressing \textbf{GQ 10.2.d.2}, which concerns the assumptions made about which candidate features are predictive of job performance, developers may document prompt templates used to query the LLM and reports of ablation tests assessing how the removal of specific CV attributes affects the generated score; when open LLMs are used, additional forms of evidence may be available, such as feature importance analyses derived directly from the model. 
Similarly, in relation to \textbf{GQ Annex IV.2.b.5}, which concerns the computation of the system’s scoring objective and the transparency of its underlying rationale, providers may document the evaluation metrics used to test the scoring mechanism (e.g., ranking accuracy, calibration, or fairness metrics) and rely on technical documentation such as model cards supplied by the third-party provider. When open models are adopted, further transparency may be achieved by detailing the scoring algorithms or intermediate computation steps that contribute to the final ranking. 

As illustrated in the examples above, the type and granularity of evidence that developers can document to substantiate compliance may vary depending on the level of transparency available for the AI components integrated into the system. 
In particular, when AI-assisted recruitment tools rely on third-party closed GPAI models, the ability of downstream providers to generate the evidence prompted by the toolkit may depend on the information made available by upstream model providers. This dependency can give rise to informational gaps along the AI value chain, an issue discussed in the next Section \ref{sec:challenges}.

\subsection{Informational gaps along the AI-value chain} 
\label{sec:challenges}
\begin{figure}[t!]
    \centering
    \includegraphics[trim=0cm 25mm 0cm 30mm, clip, width=\linewidth]{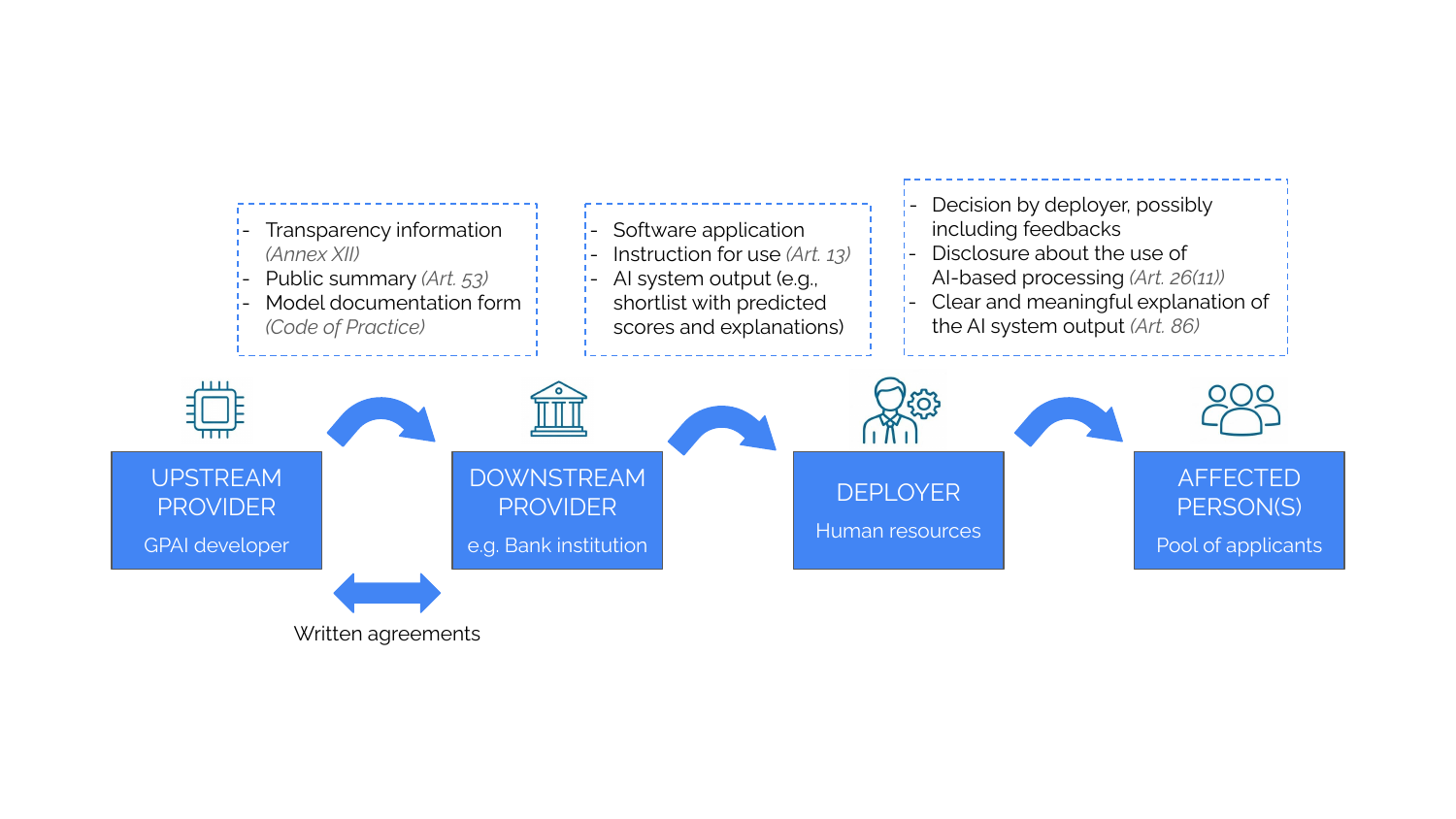}
    \caption{\textbf{Informational flow along the AI value chain in the context of high-risk AI systems integrating a third-party GPAI model.} Pursuant to the provisions of the AI Act for high-risk AI systems, the upstream provider of the GPAI model is required to make available a public summary and to disclose to the downstream provider (\textit{e.g.}, a banking institution) specific transparency information concerning the model, along with the Model Documentation Form provided by the Code of Practice (if signatory). Supplementary information may additionally be disclosed through written agreements between the parties. The downstream provider delivers the software application, together with instructions for use, to the deployer (\textit{e.g}., a human resources department), who operates the AI system and has visibility over its output, ensuring continuous human oversight of the high-risk system. On this basis, the deployer communicates decisions to the affected persons (\textit{i.e}., the pool of applicants), informing them of the role played by the AI system in the decision-making. In preparation of potential cases where an applicant wishes to exercise their right to an explanation, the deployer must be able to provide clear and meaningful details regarding the main elements of the decision taken (Article 86).}
    \label{fig:AI_Value_Chain}
\end{figure}

The guiding questions presented in the previous section presuppose that downstream providers have access to sufficient technical and organizational information to substantiate compliance with the AI Act. 
However, where high-risk systems integrate proprietary GPAI models, such access may be significantly constrained by legal and contractual limitations governing intellectual property and confidential business information or trade secrets (Article 53(1)(b)). Furthermore, GPAI model providers concentrate technical know-how, training resources (\textit{e.g}., datasets, computing infrastructures and model architectures), evaluation resources, and bargaining power in ways that downstream AI providers typically cannot match.
This structural limitation has direct accountability implications for downstream providers, who remain legally responsible for conformity assessment under Article~43. It also has indirect consequences, first for deployers and secondarily for affected persons, ultimately weakening fundamental rights protection~\parencite{Oueslati2025,Engler2022}.
Figure~\ref{fig:AI_Value_Chain} illustrates the informational flows and actor relationships along the AI value chain that underlie the challenges discussed in this section.

Against this background, the following examines the scope and limits of the transparency obligations imposed on GPAI providers under Annex XII (in contrast with the broader documentation obligations under Annex XI), as well as the role of the EU GPAI Code of Practice. 
We discuss the extent to which these instruments are capable of addressing informational asymmetries along the AI value chain and highlight the potential role of contractual clauses as a complementary mechanism to facilitate proportionate information-sharing among the actors involved.

GPAI providers are required to make available to downstream providers the documentation described in Annex XII(2), which encompasses only high-level information on the model, its development process, and data provenance. The information contained in Annex XII represents a strict subset of the broader technical documentation specified in Annex XI, which, under Article 53(1)(a), is to be provided solely to the AI Office and national competent authorities. This includes, for instance, a detailed specification of the model and training process and methodologies, along with the rationale behind key design choices and assumptions. 
Access to more detailed information, such as that described in Annex XI, would enable downstream providers to assess and implement the necessary measures to comply with the obligations set out  in Section II. 

A key European Commission initiative in this context is the \textit{Code of Practice}~\parencite{CodeOfPractice25}, developed in line with Article~56. Among multiple objectives, it seeks to facilitate transparency through a ``\textit{Model Documentation Form}'' specified in the Transparency chapter~\parencite{TransparencyCoP25}. 
While it encourages GPAI model providers to record key model details such as the intended purpose, known limitations and integration requirements, it largely adheres to Annex XI/XII 
specifications for disclosing limited information to downstream providers.
Although broader access to information by national competent authorities and the AI Office is justified by their supervisory functions, the documentation accessible to downstream providers remains comparatively limited. 
Consequently, the informational constraints affecting conformity assessment at the downstream level largely persist~\parencite{Veale2025}.
Moreover, the Code of Practice is a voluntary, not legally binding, transitional tool established via a multi-stakeholder process (including mostly self-nominated experts), in the absence of harmonised standards set by private international organisations \parencite{Veale2025}.
According to Article 53(4), GPAI providers who choose not to adopt it shall only demonstrate alternative means of compliance, thus leaving upstream providers with broad discretion over information to share that is nonetheless essential for conformity assessment. 
Additionally, while restrictions on disclosure grounded in trade secret protection may be proportionate, in practice they can further limit the technical information available to downstream providers. 
This, in turn, constrains the evidentiary basis upon which compliance of the system placed on the market can be demonstrated~\parencite{Ebers2025}.
As a result, the Code of Practice fails to address the significant information gaps left open by the AI Act that downstream providers experience. 

The AI Act does anticipate the potential emergence of difficulties in accessing necessary information and addresses this in Article 25(4) by requesting the downstream and upstream providers to bridge the information gap through written agreements aimed at ensuring the necessary information, capabilities, technical access, and other assistance. 
These agreements should reflect the approach set out in Recital 101 which guides the interpretation of Article 53. Specifically, it emphasises that downstream providers must have a \textit{``good understanding''} of the GPAI model and its capabilities, and that transparency measures should be proportionate to what is needed for downstream providers to meet their obligations under the AI Act and other applicable legal rules. Accordingly, upstream providers should clearly communicate whether and under what conditions their model is suitable for the intended use in high-risk systems, thereby reducing the risk of non-compliant or unsafe deployments (Article 25(2)).
Such legal instruments would foster shared accountability among the actors involved and giving concrete effect to the AI Act goals pursued by the co-legislators.
In this regard, to facilitate cooperation and ensure a consistent implementation of these agreements between upstream and downstream providers, Article 25(4) encourages the AI Office with developing and recommending voluntary model terms \parencite{Ebers2025}.
To date, however, these terms have not yet been published.

Some attempts in this direction have already been pursued by the Community of Practice on Public Procurement of AI with the  ``EU AI model for contractual clauses'', drafted with the support of the European Commission~\parencite{eumodelclauses2023}. The clauses contained therein constitute an emerging reference framework aimed at promoting consistent solutions and reducing negotiation asymmetries between upstream GPAI providers and downstream providers. In addition, this observation has been highlighted by other financial entities, such as the Association of Luxembourg Fund Industry \parencite{ALFI2025}. Among these, the Society for Computers \& Law (SCL) AI Group offers sample clauses to assist parties in allocating responsibilities and demonstrating compliance under the AI Act \parencite{SCLAIClauses2024}.

\section{Conclusions} 
\label{sec:conclusions}

This paper examined how compliance with the AI Act can be assessed in practice by downstream providers that integrate third-party GPAI models into high-risk applications. 
Using an LLM-based recruitment pipeline as a case study, we proposed \hrrai{}, a compliance toolkit that translates Articles 9–15 and Annex IV into guiding questions that help providers identify what evidence is necessary to demonstrate conformity across risk management, data governance, technical documentation, transparency, human oversight, and accuracy, robustness and cybersecurity.
Additionally, we discussed the informational gaps between upstream and downstream providers in the case of integration of third-party GPAI models. 
In this setting, compliance becomes contingent on value-chain information flows that are only partially guaranteed by the AI Act, with direct implications for accountability and conformity assessment. 

Our work can be further developed in several research directions.
While our analysis currently targets the core techno-normative requirements for high-risk AI systems as laid out in Chapter III, Section 2 of the AI Act, it could be strengthened by integrating other operational instruments  that provide complementary legal safeguards, such as the Fundamental Rights Impact Assessment (FRIA) required by Article 27.
The second research direction aims to further develop the participatory approach by broadening the range of participants to include not only academic researchers and internal stakeholders from the partner banking institution but also affected persons and civil society organisations. 

Additional investigations should focus on evaluating \hrrai{} through qualitative and quantitative user studies conducted in real-world settings that implement the pipeline analysed in the current study, namely AI-based recruitment systems integrating third-party 
GPAI models. 
In conclusion, further studies could examine the applicability of the \hrrai{} toolkit across scenarios that differ in terms of technology (e.g., discriminative against generative AI) and access (e.g., open against close models). 

\printbibliography
\clearpage
\newgeometry{top=1cm,left=1.5cm,bottom=1cm,right=1.5cm,includefoot}
\raggedbottom
\appendix

\section{\hrrai{} compliance toolkit}\label{app:hrrai}
\begin{table}[h]
    \centering
    \begin{tabular}{m{0.02\textwidth}m{0.98\textwidth}}
    \rowcolor[HTML]{cfe2f3}
         \multicolumn{2}{c}{
         \href{https://eur-lex.europa.eu/eli/reg/2024/1689/oj/eng\#art_9}{\large Art. 9: Risk management system}} \\ 
         & \\
         \cellcolor[HTML]{cfe2f3} \rotatebox{90}{\textbf{Interpretation guidelines}} & 
         \href{https://eur-lex.europa.eu/eli/reg/2024/1689/oj/eng#rct_65}{Recital 65}: 
         \textit{The risk-management system should consist of a continuous, iterative process that is planned and run throughout the entire lifecycle of a high-risk AI system. 
         [...] This process should ensure that the provider identifies risks or adverse impacts and implements mitigation measures for the known and reasonably foreseeable risks of AI systems to the health, safety and fundamental rights in light of their intended purpose and reasonably foreseeable misuse, including the possible risks arising from the interaction between the AI system and the environment within which it operates. 
         The risk-management system should adopt the most appropriate risk-management measures in light of the state of the art in AI. When identifying the most appropriate risk-management measures, the provider should document and explain the choices made and, when relevant, involve experts and external stakeholders. 
         In identifying the reasonably foreseeable misuse of high-risk AI systems, the provider should cover uses of AI systems which
         [...] may nevertheless be reasonably expected to result from readily predictable human behaviour in the context of the specific characteristics and use of a particular AI system. 
         Any known or foreseeable circumstances related to the use of the high-risk AI system in accordance with its intended purpose or under conditions of reasonably foreseeable misuse, which may lead to risks to the health and safety or fundamental rights should be included in the instructions for use that are provided by the provider. This is to ensure that the deployer is aware and takes them into account when using the high-risk AI system. 
         [...] The providers however are encouraged to consider such additional training measures to mitigate reasonable foreseeable misuses as necessary and appropriate.} 
         \\ 
    \end{tabular}
\end{table}

\begin{longtable}{p{0.4\textwidth}p{0.6\textwidth}}
\rowcolor[HTML]{cfe2f3}\textbf{Compliance checklist} & \textbf{Guiding questions} \\
& \\
\endhead
\multirow{3}{0.4\textwidth}{\parbox{0.4\textwidth}{1. A \textbf{risk management system} is established, implemented, documented and maintained in relation to high-risk AI systems.}} 
&9.1.1 Has a risk management system (RMS) been established?\\
&9.1.2 How is the RMS implemented?\\
&9.1.3 How is the RMS documented and maintained?\\
\midrule\multirow{6}{0.4\textwidth}{\parbox{0.4\textwidth}{2. The risk management system is understood as a \textbf{continuous iterative process} planned and run throughout the entire lifecycle of an high-risk AI system, and a systematic review and updating process is defined.}} 
&9.2.1 How do you define a “continuous and iterative process” in your context?\\
&9.2.2 What rules or conditions trigger a new RMS iteration?\\
&9.2.3 For each RMS iteration, what is the cadence, and how do you justify that it is effective in reducing identified risks?\\
&9.2.4 Is the RMS applied throughout the entire AI system lifecycle? How does it differ across lifecycle phases?\\
&9.2.5 How is the systematic review of the RMS defined?\\
&9.2.6 How is the RMS updating process defined? \\
\midrule\multirow{4}{0.4\textwidth}{\parbox{0.4\textwidth}{2(a). The risk management system identifies and analyses the known and the reasonably foreseeable risks that the high-risk AI system can pose to health, safety or fundamental rights when the high-risk AI system is used in accordance with its intended purpose.}} 
&9.2.a.1 What known risks to health, safety, or fundamental rights may arise when the AI system is used as intended?\\
&9.2.a.2 What reasonably foreseeable risks to health, safety, or fundamental rights may arise when the AI system is used as intended?\\
& 9.2.a.3 Have you identified and analysed the risks to health, safety, or fundamental rights that may arise from integrating third-party GPAI model(s) into the recruitment system (e.g., in CV parsing, candidate scoring, or ranking), including risks related to limited transparency, model updates outside the provider’s control, and potential biases in model outputs?\\
&9.2.a.4 Which fundamental rights are most likely to be affected by the system? How does your Fundamental Rights Impact Assessment (Art. 27) identify, measure, and mitigate these impacts?\\
&9.2.a.5 Which methods and evidence do you use to analyse known and reasonably foreseeable risks? For each risk, provide relevant metrics, thresholds, documented findings, and the mitigation or acceptance decision. \\
\midrule\multirow{4}{0.4\textwidth}{\parbox{0.4\textwidth}{2(b). The risk management system estimates and evaluates the risks that may emerge when the AI system is used in accordance with its \textbf{intended purpose and under condition of reasonable and foreseeable misuse}.}} 
&9.2.b.1 Which methods and/or tools does the RMS use to estimate and evaluate risks? Please provide details and sample evidence.\\
&9.2.b.2 How do you validate the robustness of risk estimates and evaluations? Describe the validation methods used.\\
&9.2.b.3 List the reasonably foreseeable misuse scenarios identified for this system. For each scenario, indicate the likelihood, potential harms, and how and where it is addressed in the system pipeline.\\
&9.2.b.4 Why are the chosen risk estimation and evaluation methods appropriate for the system’s intended purpose? \\
\midrule\multirow{5}{0.4\textwidth}{\parbox{0.4\textwidth}{2(c). The risk management system evaluates the (other) risks possibly arising from the analysis of the \textbf{data gathered by the post-market monitoring system} in compliance with Article 72.}} 
&9.2.c.1 How does post-market monitoring data feed into the RMS?\\
&9.2.c.2 Is a documented integration pipeline in place between post-market monitoring and the RMS? If so, provide its specification.\\
&9.2.c.3 Which analytical methods and criteria are applied to post-market monitoring data to detect unforeseen risks?\\
&9.2.c.4 Is a corrective process defined for each newly identified risk detected through post-market monitoring?\\
&9.2.c.5 If pipeline updates are planned, how will they be implemented? \\
& 9.5.c.6 What information about the third-party GPAI model (e.g., model cards, technical documentation, usage constraints, evaluation results) is available and incorporated into your risk management process?\\
&9.5.c.7 How do you document limitations, assumptions, or uncertainties related to the third-party GPAI components to enable effective oversight and mitigation of risks?\\
\midrule
2(d). The risk management system adopts appropriate and targeted risk management measures pursuant point 2(a). &9.2.d.1 Which measures have you adopted to ensure an appropriate and targeted assessment of risks to health, safety, and fundamental rights? \\
\midrule\multirow{2}{0.4\textwidth}{\parbox{0.4\textwidth}{3. The risk management system addresses only the risks that can be reasonably mitigated or eliminated by appropriate development or design of the AI system, or by the provision of adequate technical documentation.}} 
&9.3.1 Have you identified the risks that can be reasonably mitigated or eliminated? Provide a list of such risks and describe the methods and criteria used to determine whether a particular risk is reasonably mitigated.\\
&9.3.2 Are there foreseeable risks to health, safety, or fundamental rights that cannot reasonably be mitigated or eliminated? If so, explain why and how this determination was made. \\
\midrule\multirow{5}{0.4\textwidth}{\parbox{0.4\textwidth}{4. The adopted risk management measures (set out in Article 9.2(d)) gives consideration to the effects and possible interactions resulting from their \textbf{combined application with the requirements} set out in Section 2, with a view to minimising risks more effectively while achieving an appropriate balance in implementing the measures to fulfil those requirements.}} 
&9.4.1 Have you identified interactions between the RMS and other Section 2 requirements?\\
&9.4.2 How is the RMS designed to support compliance with all Section 2 requirements, including in cases of system modification?\\
&9.4.3 How do you balance RMS implementation with other Section 2 requirements?\\
&9.4.4 How do you justify trade-offs between RMS measures and other Section 2 requirements?\\
&9.4.5 What inherent or foreseeable risks arise from your chosen trade-off strategy? \\
\midrule\multirow{4}{0.4\textwidth}{\parbox{0.4\textwidth}{5. The risk management measures (Article 9.2(d)) are so that the \textbf{residual risk} associated with each hazard, as well as the overall residual risk of the AI system, is judged to be acceptable.}} 
&9.5.1 Which hazards are present in your specific application?\\
&9.5.2 How do you define “acceptable residual risk”? How is it measured?\\
&9.5.3 What are the residual risks associated with each hazard? Why are they considered acceptable?\\
&9.5.4 What is the overall residual risk of the AI system? Why is it considered acceptable? \\\midrule
\multicolumn{2}{l}{\cellcolor[HTML]{efefef}\textit{The \textbf{risk management measures} shall ensure:}} \\
\multirow{3}{0.4\textwidth}{\parbox{0.4\textwidth}{5(a). Elimination or reduction of risks identified as far as technically feasible by adequate design and development of the High-Risk AI system.}} &9.5.a.1 What technical design and development strategies have you implemented to eliminate or reduce the risks identified under Article 9.2?\\
&9.5.a.2 How do these solutions eliminate or reduce the identified risks? Which alternatives were considered, and why was the chosen approach preferred?\\
&9.5.a.3 What technical limitations, if any, does your solution present?\\
\midrule\multirow{2}{0.4\textwidth}{\parbox{0.4\textwidth}{5(b). (Where appropriate) Implementation of adequate mitigation and control measures addressing risks that cannot be eliminated.}}
&9.5.b.1 Why can certain risks not be eliminated? For each such risk, provide an analysis explaining why elimination is not reasonably feasible.\\
&9.5.b.2 What mitigation and control measures have been implemented for risks that cannot be eliminated, and why were these measures chosen? \\
\midrule\multirow{5}{0.4\textwidth}{\parbox{0.4\textwidth}{5(c). Provision of information (following Article 13) and, where appropriate, training to the deployer so as to eliminate or reduce risks related to the use of the high-risk AI system taking into consideration technical knowledge, experience, education, expected training by the deployer, and the presumable context in which the system is intended to be used.}} 
&9.5.c.1 What information required under Article 13 is provided to deployers to eliminate or reduce risks?\\
&9.5.c.2 What level of AI literacy (Art. 4), including technical knowledge, experience, education, and training, is required from deployers to reduce or eliminate risks?\\
&9.5.c.3 How does this AI literacy requirement account for the specific context of use, particularly regarding the persons or groups on whom the system is used?\\
&9.5.c.4 Is application-specific training appropriate for deployers of this system? Why or why not?\\
&9.5.c.5 What minimum level of AI literacy do you expect deployers to have? \\
&9.5.c.6 What information about the third-party GPAI model (e.g., model cards, technical documentation, usage constraints, evaluation results) is available and incorporated into your risk management process?\\
&9.5.c.7 How do you document limitations, assumptions, or uncertainties related to the third-party GPAI component to enable effective oversight and mitigation of risks?\\
\midrule\multirow{3}{0.4\textwidth}{\parbox{0.4\textwidth}{6. The High-risk AI system is \textbf{tested} for the purpose of identifying the most appropriate and targeted risk management measures, ensuring consistent performance for its intended purpose, and is in compliance with the requirements set out in Section 2.}} 
&9.6.1 Did you verify that testing is conducted under conditions representative of the production environment and that the system performs consistently for its intended purpose?\\
&9.6.2 If not, what is the risk that the selected risk management measures may fail to protect health, safety, or fundamental rights?\\
&9.6.3 How do you determine the most appropriate and targeted risk management measures? Which alternatives were considered, and why was the chosen option selected? \\
\midrule\multirow{3}{0.4\textwidth}{\parbox{0.4\textwidth}{7. \textit{(Optional)} The High-risk AI system is tested for the purpose of identifying the most appropriate and targeted risk management measures, ensuring consistent performance for its intended purpose, in \textbf{real-world conditions} outside AI regulatory Sandboxes (Art. 60).}} 
&9.7.1 Did you conduct testing in real-world conditions outside an AI regulatory sandbox? If so, did you comply with Article 60 requirements?\\
&9.7.2 If not, why was real-world testing not performed, and what alternative validation measures were taken?\\ \\
\\
\midrule\multirow{7}{0.4\textwidth}{\parbox{0.4\textwidth}{8. The testing of high-risk AI systems is performed as appropriate, at any time throughout the development process, and, in any event, \textbf{before deployment}. Testing is carried out using prior defined appropriate metrics for the intended purpose.}} 
&9.8.1 Which metrics and probabilistic thresholds have been defined for the system’s intended purpose, and why are they appropriate?\newline
- Which alternatives were considered and how did experimental comparisons inform the choice? \newline
- What are the limitations and foreseeable risks associated with the chosen metrics and thresholds? \\
&9.8.2 At which stages of development was testing performed? \\
&9.8.3 Was the system tested prior to market or service release? Please provide supporting evidence.\\
&9.8.4 How do you ensure that predefined metrics and thresholds remain appropriate under real-world conditions and in the intended context of use?\\
&9.8.5 How will metrics and thresholds be adapted to real-world scenarios and evolving data conditions (e.g., unexpected data features or new feature types)? \\
\midrule\multirow{2}{0.4\textwidth}{\parbox{0.4\textwidth}{9. During the High-risk AI system implementation, providers considered whether or not it is likely to have a \textbf{negative impact on persons under the age of 18 or other vulnerable groups} in view of the system intended purpose.}}
&9.9.1 Are there foreseeable adverse impacts on persons under 18 or other vulnerable groups arising from the use of this system?\\
&9.9.2 If so, what are these impacts, and how does the RMS mitigate them? \\ \\ \\
\midrule\multirow{2}{0.4\textwidth}{\parbox{0.4\textwidth}{10. The aspects provided in paragraphs 1 to 9 are a part of, or combined with, the risk management procedures established under \textbf{other relevant provisions of Union law}\footnote{Intended for providers of high-risk AI systems that are subject to requirements regarding internal risk management processes under other relevant provisions of Union law.}}.
} &9.10.1 Which other Union law provisions imposing internal risk management obligations apply to this system?\\
&9.10.2 How are those internal risk management procedures integrated with the RMS established under the AI Act? \\
\bottomrule 

\end{longtable}

    \par\mbox{}\clearpage
    \begin{table}[h]
    \centering
    \begin{tabular}{m{0.02\textwidth}m{0.98\textwidth}}
    \rowcolor[HTML]{cfe2f3}
         \multicolumn{2}{c}{
         \href{https://eur-lex.europa.eu/eli/reg/2024/1689/oj/eng\#art_10}{\large Art. 10: Data and data governance}} \\ 
         & \\
         \cellcolor[HTML]{cfe2f3} \rotatebox{90}{\textbf{Interpretation guidelines}} & 
         \href{https://eur-lex.europa.eu/eli/reg/2024/1689/oj/eng#rct_67}{Recital 67}: 
         \textit{High-quality data [...] plays a vital role in [...] ensuring the performance of many AI systems [...]. Data sets for training, validation and testing, including the labels, should be relevant, sufficiently representative, and to the best extent possible free of errors and complete in view of the intended purpose of the system. [...] The data sets should also have the appropriate statistical properties, including as regards the persons or groups of persons in relation to whom the high-risk AI system is intended to be used, with specific attention to the mitigation of possible biases in the data sets, that are likely to affect the health and safety of persons, have a negative impact on fundamental rights or lead to discrimination prohibited under Union law, especially where data outputs influence inputs for future operations (feedback loops). [...] Data sets should take into account, to the extent required by their intended purpose, the features, characteristics or elements that are particular to the specific geographical, contextual, behavioral or functional setting which the AI system is intended to be used. [...]}\newline

         \href{https://eur-lex.europa.eu/eli/reg/2024/1689/oj/eng#rct_68}{Recital 68}: 
         \textit{For the development and assessment of high-risk AI systems, certain actors such as providers, notified bodies and other relevant entities, [...] should be able to access and use high-quality data sets [...] for the training, validation and testing of AI systems. [...]}\newline

         \href{https://eur-lex.europa.eu/eli/reg/2024/1689/oj/eng#rct_69}{Recital 69}: 
        \textit{The right to privacy and to protection of personal data must be guaranteed throughout the entire lifecycle of the AI system. [...] The principles of data minimisation and data protection by design and by default, as set out in Union data protection law, are applicable when personal data are processed. Measures [...] may include not only anonymisation and encryption, but also the use of technology that permits algorithms to be brought to the data and allows training of AI systems without the transmission between parties or copying of the raw or structured data themselves [...].}\newline


         \href{https://eur-lex.europa.eu/eli/reg/2024/1689/oj/eng#rct_70}{Recital 70}: 
        \textit{[...] Providers should, exceptionally, to the extent that it is strictly necessary for the purpose of ensuring bias detection and correction in relation to the high-risk AI systems, subject to appropriate safeguards for the fundamental rights and freedoms of natural persons [...] be able to process also special categories of personal data, as a matter of substantial public interest [...].}
                 
    \end{tabular}
\end{table}

\begin{longtable}{p{0.4\textwidth}p{0.6\textwidth}}
\rowcolor[HTML]{cfe2f3}\textbf{Compliance checklist} & \textbf{Guiding questions} \\
& \\
\endhead
\multicolumn{2}{l}{
\cellcolor[HTML]{efefef}
\parbox{\linewidth}{
\textit{
2. Data governance and management practices should be appropriate for the intended purpose of the high-risk AI system and concern:
}}
} \\
\multirow{10}{0.4\textwidth}{\parbox{0.4\textwidth}{2(a). \textbf{Design choices for} the \textit{testing data}.}} & 10.2.a.1 Which format did you use for CVs and job descriptions (e.g., structured templates vs. unstructured formats like PDFs)? \\
&10.2.a.2 What assumptions about data quality or availability have been made? \\
&10.2.a.3 Does the data pre-processing pipeline include parsing of unstructured input data? If so, how was it conducted? \\
&10.2.a.4   Which languages are supported for CVs and job descriptions? If applicable, how does translation occur, and how did you verify its quality? \\
&10.2.a.5 In the case of unstructured input data, is the output of the parsing pipeline structured according to a homogeneous template? What information is extracted and why? \\
&10.2.a.6  Why did you choose to use a public and/or private test dataset? \\
&10.2.a.7  Is the test data you chose representative of the real context of use of the system? \\
&10.2.a.8 How did you choose the target labels for testing? \\
&10.2.a.9 How did you choose the evaluation metrics for testing? \\
&10.2.a.10 How did you choose to annotate your testing data, if needed? If humans were involved: \\
&-     How was the selection of human annotators conducted (e.g., in terms of AI literacy and domain expertise)? \\
&-     Do human annotators have an internal or external function within the organization? \\
&-     Did you design the annotation process to ensure the inclusion of a variety of perspectives? \\
\midrule\multirow{7}{0.4\textwidth}{\parbox{0.4\textwidth}{2(b). \textbf{Sources and collection processes for} the \textit{testing data}, including details on the original purpose, in case personal data is involved.}} &10.2.b.1 Are the sources of the testing data comprehensively documented in either internal or external documents? \\
&10.2.b.2   How does the collection process handle data covered by copyright? Is the final collected data protected by copyright? \\
&10.2.b.3  How does the collection process handle potentially illegal (e.g., offensive or non-consensual) data? Does the final collected data contain potentially illegal data? \\
&10.2.b.4  Are the data sources for testing data all public? If not, which private data sources did you collect data from? \\
&10.2.b.5  Do the data sources for testing data contain personal information? If so, did you record their original purpose? Moreover, if special categories of personal data are included, did you comply with Article 10(5)? \\
&10.2.b.6  How is test data tracked or updated? In the case of an update, please provide a detailed account of the sources of additional test data (e.g., feedback loops) and the collection processes. \\
&10.2.b.7  Did you record the data collection procedure, including the computational methods employed for extraction (e.g., discriminative or generative methods)? \\
\midrule\multirow{8}{0.4\textwidth}{\parbox{0.4\textwidth}{2(c). \textbf{Pre-processing steps} taken to clean, annotate, label, update, enrich and aggregate the \textit{testing data}.}} &10.2.c.1 Was data cleaning carried out during the preprocessing of the testing data? If so, please provide a detailed account of the procedures that were implemented. \\
&10.2.c.2  Was data annotation carried out during the pre-processing of the testing data? If so, please provide a detailed account of the procedures that were implemented. In particular, how were candidate scores generated for testing? \\
&10.2.c.3   Was data annotation conducted through automated tasks and/or humans? How was the annotation process carried out? \\
&10.2.c.4 If third-party AI or GPAI components are used in data preparation operations (e.g., annotation, parsing, enrichment or representation generation), what documentation has been obtained from the third-party provider regarding how these components transform the data?\\
&10.2.c.5 How have you assessed whether third-party AI or GPAI components used in data preparation may introduce errors or representational distortions affecting the quality and completeness of the testing datasets?\\
&10.2.c.6 To ensure the inclusion of a variety of perspectives, were there multiple annotators involved? How did you account for their potential disagreement? \\
&10.2.c.7 Was testing data enriched during pre-processing (e.g., with synthetic data)? If so, please provide a detailed account of the procedures that were implemented. \\
&10.2.c.8  Was testing data aggregation carried out during the pre-processing of the testing data? If so, please provide a detailed account of the procedures that were implemented for aggregating both input data and labels. \\
&10.2.c.9   Is test data updated periodically? If so, please provide a detailed account of the preprocessing steps that were implemented to update the test data, including strategies for handling inconsistency with respect to historical test data. \\
\midrule\multirow{10}{0.4\textwidth}{\parbox{0.4\textwidth}{2(d). The formulation of \textbf{assumptions}, in particular with respect to the information that the testing data are supposed to measure and represent.}} &10.2.d.1 Which elements of the candidate's CV do you treat as predictors of future job performance, and are those features included in your testing dataset? \\
&10.2.d.2  What explicit assumptions are you making about which candidate features (e.g., education, work history, test results) are predictive of job performance? \\
&10.2.d.3  Do you assume equivalence between similar qualifications across different geographies, education systems, or socio-economic backgrounds? \\
&10.2.d.4  Do the datasets measure competencies directly relevant to the role or are proxies being used? \\
&10.2.d.5  Are you making assumptions about cultural fit or personality traits that may not be objectively measurable? \\
&10.2.d.6  Are assumptions about ``successful'' past candidates potentially perpetuating historical under-representation or bias? \\
&10.2.d.7  Do the assumptions in the testing data hold across different contexts, such as job functions or geographical contexts? \\
&10.2.d.8  Are assumptions regularly revisited and updated as the nature of work and job requirements evolve? \\
&10.2.d.9 What is the expected pool of applicants (e.g., in terms of demographics, geographies, educational qualification systems, or socio-economic backgrounds)? \\
&10.2.d.10 How did you ensure that the target label represents the intended purpose of the system? Provide a detailed account taking into consideration all intermediate steps of the pipeline, in particular the output of the candidate scoring step. \\
\midrule\multirow{6}{0.4\textwidth}{\parbox{0.4\textwidth}{2(e). Data governance and management practices should be appropriate for the intended purpose of the high-risk AI system and concern: \textbf{The assessment of availability, quantity and suitability} of the testing data.}} &10.2.e.1 Although you opted for a pipeline based on an LLM with no fine-tuning, did you also assess the availability, quantity, and suitability of data in view of implementing an alternative pipeline that instead supports the possibility of conducting training and fine-tuning other than testing? If so, please provide documentation and justification for this assessment and decision. \\
&10.2.e.2   Could you find data that was compatible with your context of use to test the model? What strategies did you employ to address the potential lack of data? \\
&10.2.e.3 In order to accurately represent and measure the task(s) the model is intended to carry out within the context of use, how did you guarantee that all information contained in the test data was comprehensive and sufficient? \\
&10.2.e.4   Is the size of the test sample large enough to guarantee statistical significance of the assessments, including when stratified by features of interest (e.g., assessment of fairness metrics)? \\
&10.2.e.5    Is the testing dataset representative of the diversity of applicants expected in deployment (gender, age, disability, ethnicity, educational background, etc.) for each job application option? \\
&10.2.e.6  Could you identify persons or groups of persons at risk of being under-represented in the testing data? If so, how is this addressed? \\
&10.2.e.7 If third-party AI or GPAI components are used in data preparation operations, how have you verified that they do not negatively affect the representativeness of the testing datasets?\\
\midrule\multirow{7}{0.4\textwidth}{\parbox{0.4\textwidth}{2(f). An examination of possible\textbf{ biases} that are likely to affect the health and safety of persons, have a negative impact on fundamental rights or lead to discrimination prohibited under Union law, especially where data outputs influence inputs for future operations.}} &10.2.f.1  According to which legislation (e.g., the Charter of Fundamental Rights, EU and national legislation) did you define biases in your application? \\
&10.2.f.2   Have potential sources of bias been identified in the test dataset? For instance: \\
&-     Does input data contain sensitive attributes (e.g., gender, ethnicity, disability, age, or socio-economic background)? \\
&-     Are the outputs of the AI system (e.g., shortlisted candidates) used as inputs for future training or testing datasets? If yes, could this feedback loop reinforce historical under-representation or biased patterns (e.g., fewer women in STEM roles leading to fewer women being shortlisted in the future)? \\
&-     are the testing procedures sufficient to identify biases derived from upstream and/or external data sources that feed the AI system? \\
&10.2.f.3  In which way may the identified biased AI outputs affect downstream processes (e.g., interview invitations, job offers)? \\
&10.2.f.4  Could proxy variables (e.g., postal code, education institution, or employment gaps) indirectly encode protected characteristics? \\
&10.2.f.5  Are there risks that automated rejection patterns disproportionately affect vulnerable groups, leading to social or economic harm? \\
&10.2.f.6  Are performance metrics (e.g., job success indicators) appropriate for the specific context of use? \\
&10.2.f.7  How is transparency ensured so that the deployers (i.e., recruiters) understand the assumptions and risks in the use of the AI system? \\
&10.2.f.8 If third-party AI or GPAI components are used in data preparation (e.g., parsing, annotation, enrichment), have you assessed whether they may introduce biases that could affect fundamental rights or lead to discrimination?\\
\midrule
\multirow{14}{0.4\textwidth}{\parbox{0.4\textwidth}{2(g). Appropriate measures to \textbf{detect, prevent and mitigate such biases} are put in place (if identified in (f)).}} &10.2.g.1   Which techniques and processes did you put in place to detect bias before deployment? \\
&10.2.g.2  Which bias detection techniques (e.g., statistical parity checks and adverse impact analysis) have been applied to the testing datasets? For instance, did you assess for imbalances in representation of groups across sensitive attributes and target-related stratification (e.g., job description or level)? \\
&10.2.g.3  Did you perform bias testing across the full recruitment funnel (parsing, scoring, screening)? \\
&10.2.g.4  How are proxy variables (e.g., postal code, education institution, or employment gaps) detected and how is the bias deriving from them prevented and mitigated? \\
&10.2.g.5  How frequently is bias reassessed as data or models are updated? \\
&10.2.g.6  Which techniques and processes did you put in place to prevent bias? \\
&10.2.g.7 In order to prevent biases, were steps taken during data collection to ensure that protected groups were adequately represented in the testing datasets? \\
&10.2.g.8  Did you verify that data pre-processing steps (e.g., normalization, feature engineering) don't amplify existing disparities? \\
&10.2.g.9 Are domain experts in HR or ethics involved in reviewing design choices and non-discriminatory assumptions? \\
&10.2.g.10 If bias is identified, what corrective steps (e.g., rebalancing, re-weighting, human oversight, abstention mechanism) are taken before deployment? \\
&10.2.g.11  How do you ensure that bias mitigation does not negatively affect model performance or robustness of predictions? \\
&10.2.g.12  How did you mitigate biases arising from feedback loops (e.g., where biased outputs, such as favoring certain profiles, influence future data collection)? \\
&10.2.g.13  Are rejected or under-represented candidates systematically excluded from future data? How did you prevent and mitigate the risks that automated rejection patterns disproportionately affect vulnerable groups? \\
&10.2.g.14 Are there mechanisms to audit long-term effects of AI-driven recruitment on workforce diversity and equal opportunity? \\
\midrule\multirow{6}{0.4\textwidth}{\parbox{0.4\textwidth}{2(h). The identification of relevant \textbf{data gaps or shortcomings} preventing compliance with AI Act are identified and addressed.}} &10.2.h.1 What mechanisms are in place to systematically identify missing, incomplete, or low-quality data in the testing datasets? \\
&10.2.h.2 Are there particular job categories, applicant demographics (protected groups in particular), or stages in the recruitment pipeline where data is sparse or inconsistent? \\
&10.2.h.3 Are there variables or indicators crucial for fairness assessment (e.g., socio-economic background, accessibility needs) that are not currently collected or are unavailable? \\
&10.2.h.4 How could existing data gaps or quality issues affect the system's accuracy, robustness, fairness, transparency and auditability of recruitment outcomes? \\
&10.2.h.5 What processes exist for addressing identified data gaps (e.g., data augmentation, collection of new representative samples, re-weighting)? What are the possible limitations of these processes? \\
&10.2.h.6   How does the system handle out-of-distribution samples (after deployment)? \\
\midrule
\pagebreak
\multirow{2}{0.4\textwidth}{\parbox{0.4\textwidth}{3. The testing data is \textbf{relevant, sufficiently representative}, and to the best extent possible, \textbf{free of errors and complete} in view of the intended purpose. [They shall have the appropriate statistical properties, including, where applicable, as regards the persons or groups of persons in relation to whom the high-risk AI system is intended to be used. Those characteristics of the data sets may be met at the level of individual data sets or at the level of a combination thereof.]}} &\cellcolor[HTML]{efefef}
\parbox{\linewidth}{
\textit{
 For suitability of testing data, refer also to the guiding questions related to Article 10 (2)(e). 
}} \\
&10.3.1 What quality control measures are in place to detect and correct errors, duplicates, or inconsistencies in the data? \\
&10.3.2 Is the information included in the testing dataset (e.g., candidate profiles, assessment formats) sufficiently representative, error-free, and complete to reproduce the recruitment tasks or decisions the AI system is designed to support (e.g., screening, scoring, ranking)? If so, justify why. \\
&10.3.3 Are all data attributes included in the testing dataset directly relevant to job performance or role requirements? Or are there confounding, irrelevant, inconsistent, or outdated variables (e.g., university prestige, which is a typical proxy for economic background) that could introduce bias or noise? \\
&10.3.4 Does the testing dataset include sufficient diversity across protected characteristics (gender, age, ethnicity, disability, nationality, etc.) and job categories to detect differential performance or bias? How do you statistically measure representativeness? \\
&10.3.5 Is the representation of different applicant groups proportionate to the actual or expected applicant pool? How did you statistically measure representativeness? \\
&10.3.6 Are there missing values or incomplete candidate profiles, and how are they handled (e.g., imputation, exclusion, flagging)? How do you identify missing information? \\
&10.3.7 Are statistical tests performed to ensure that model performance metrics (e.g., accuracy, false positives/negatives) are not significantly different across protected attributes and job categories? \\
&10.3.8 If combining multiple datasets (e.g., test data + real-world recruitment data), how are differences in data quality, structure, and sampling handled? \\
&10.3.9 If you use real-world historical data, how are human errors, subjectivity or changes in recruitment requirements in past recruitment evaluations (e.g., human labels for the candidate scoring task) accounted for or corrected in the testing data? \\
&10.3.10 How often is the representativeness and quality of the testing data reviewed or updated? Are there procedures to detect data drift or performance degradation over time? \\
\midrule\multirow{9}{0.4\textwidth}{\parbox{0.4\textwidth}{4. The testing datasets \textbf{takes into account}, to the extent required by the intended purpose, the characteristics or elements that are particular to the \textbf{specific geographical, contextual, behavioural or functional setting} within which the high-risk AI system is intended to be used.}} &10.4.1 Does the testing dataset reflect the geographical region(s) where the recruitment AI system will be used (e.g., EU country, language, educational qualification system)? \\
&10.4.2 Are differences in education systems, qualifications, or job titles between geographical regions represented appropriately to grant robust predictions? \\
&10.4.3 If input data is not standardized, how are contextual differences in data format (e.g., CV language and structure) accounted for and possibly harmonized? \\
&10.4.4 Does the testing dataset capture the specific recruitment context and recruitment criteria (e.g., corporate recruitment vs public sector recruitment, recruitment of special employment categories)? \\
&10.4.5 Does the testing dataset reflect realistic applicant pools and job types for the intended context of use? \\
&10.4.6 Does the dataset account for differences in how candidates present themselves (e.g., communication styles, self-promotion tendencies, or the use of GPAI tools)? Is the AI system capable of handling them to avoid distortion in interpreting input data? \\
&10.4.7 How often are datasets reviewed to ensure continued relevance to the geographical, contextual and behavioral settings during production (e.g., after market or organizational changes)? \\
&10.4.8 Are there mechanisms for human oversight to identify when contextual drift (e.g., new recruitment practices or regional expansion) invalidates previous data assumptions? Which actions would be triggered? \\
&10.4.9    Does the AI system pipeline reflect the procedures typically conducted to accomplish the intended goal in the specific context of use? \\
\midrule
\multicolumn{2}{l}{
    \cellcolor[HTML]{efefef}
    \parbox{\dimexpr\textwidth-2\tabcolsep\relax}{
        \vspace{0.5em}
        \textit{5. In order to comply with Article 10 (2)(e)-(f) the provider may exceptionally process special categories of personal data, subject to appropriate safeguards for the fundamental rights and freedoms of natural persons. In addition to the provisions set out in Regulations (EU) 2016/679 and (EU) 2018/1725 and Directive (EU) 2016/680, all the following conditions must be met in order for such processing to occur:}
        \vspace{0.5em}
    }
} \\
\multirow{2}{0.4\textwidth}{\parbox{0.4\textwidth}{5(a). The bias detection and correction \textbf{cannot be effectively fulfilled} by processing other data, including synthetic or anonymised data.}} &10.5.a.1  Have you clearly identified what forms of bias (e.g., gender, ethnicity, disability) the recruitment AI system aims to detect and correct (see also Article 10(2)(e))? \\
&10.5.a.2  In order to fulfill bias detection and correction as per Article 10(2)(f-g), did you use anonymised or synthetic data? If not, justify the need for processing special categories of personal data. \\
\midrule\multirow{5}{0.4\textwidth}{\parbox{0.4\textwidth}{5(b). The special categories of personal data are subject to \textbf{technical limitations on the re-use} of the personal data, and \textbf{state-of-the-art security and privacy-preserving measures}, including pseudonymisation.}} &10.5.b.1 Have you implemented technical safeguards, such as pseudonymisation, encryption, or secure data environments, to protect special categories of personal data during processing? \\
&10.5.b.2  Have you assessed whether pseudonymisation sufficiently reduces the risk of re-identification within the recruitment context (e.g., small applicant pools)? \\
&10.5.b.3 Have you documented and enforced purpose limitation rules preventing the re-use of special categories of personal data for other functions? \\
&10.5.b.4  Are logs or audit trails in place to record who accessed special categories of personal data, when, and for what purpose? \\
&10.5.b.5  Have you verified that data storage and processing environments meet state-of-the-art security standards (e.g., encrypted storage, network isolation)? \\
\midrule\multirow{4}{0.4\textwidth}{\parbox{0.4\textwidth}{5(c). The special categories of personal data are subject to measures to ensure that the personal data processed are \textbf{secured, protected, subject to suitable safeguards}, including strict controls and documentation of the access, to avoid misuse and ensure that only \textbf{authorised persons} have access to those personal data with appropriate confidentiality obligations.}} &10.5.c.1 Have you implemented access controls (e.g., logs, monitoring, auditing mechanisms, access restrictions, role-based permissions), to ensure that only authorised personnel can access special categories of personal data? \\
&10.5.c.2  Are confidentiality obligations formally in place for all staff or third parties who have access to these data (e.g., through contracts or NDAs)? \\
&10.5.c.3  Are data protection environments for bias detection equally compliant with points 5.c.1 and 5.c.2 and well separated from the main operational recruitment system to reduce security issues? \\
&10.5.c.4 Do you ensure that any third-party tools or service providers involved in testing or bias correction comply with equivalent security and confidentiality standards? \\
\midrule\multirow{5}{0.4\textwidth}{\parbox{0.4\textwidth}{5(d). The special categories of personal data are \textbf{not to be transmitted, transferred or otherwise accessed} by other parties.}} &10.5.d.1  Have you clearly defined which entities and personnel are authorised to access special categories of personal data used for bias detection in recruitment? \\
&10.5.d.2  Have you implemented technical controls (e.g., role-based access, network restrictions, data encryption), business agreements and/or internal policies to ensure such data cannot be transmitted or shared outside authorised environments or with third parties (e.g., external vendors, auditors, cloud providers)? \\
&10.5.d.3  Have you verified that cloud storage, data hosting, or model testing environments do not allow external access or external data transfer of special categories of personal data? \\
&10.5.d.4  Are data export, download, or sharing functionalities disabled or restricted for datasets containing sensitive demographic attributes? \\
&10.5.d.5  Do system logs or audit trails record all instances of access to special categories of personal data, enabling detection of any unauthorised viewing or transmission? \\
\midrule
\multirow{5}{0.4\textwidth}{\parbox{0.4\textwidth}{5(e). The special categories of personal data are \textbf{deleted} once the bias has been corrected or the personal data has reached the end of its retention period, whichever comes first.}} &10.5.e.1 Have you defined a clear retention period for special categories of personal data used in bias detection? \\
&10.5.e.2  Do you have a documented procedure specifying when and how such data must be deleted once bias has been corrected? \\
&10.5.e.3  Is the responsibility for the deletion process clearly described, assigned, and documented? \\
&10.5.e.4  If the deletion process is automated, does this automation have clearly described technical guarantees? \\
&10.5.e.5  Are deletion events logged and verifiable through audit records and logs? \\
\midrule\multirow{3}{0.4\textwidth}{\parbox{0.4\textwidth}{5(f). The records of processing activities pursuant to Regulations (EU) 2016/679 and (EU) 2018/1725 and Directive (EU) 2016/680 include the reasons why the processing of special categories of personal data was \textbf{strictly necessary} to detect and correct biases, and why that objective could not be achieved by processing other data.}} &10.5.f.1 Have you documented in your records of processing activities the exact bias detection or correction purpose for which special categories of personal data were processed (e.g., checking differential model accuracy across gender or ethnicity)? \\
&10.5.f.2  Have you clearly justified why other data types (such as anonymised or synthetic data) could not achieve equivalent bias detection or correction results? \\
&10.5.f.3  Have you recorded how ``strict necessity'' was assessed, for instance, through a technical or statistical analysis showing that bias could not be reliably detected without such data? \\
\bottomrule 

\end{longtable}
    \par\mbox{}\clearpage
    \begin{table}[h]
    \centering
    \begin{tabular}{m{0.02\textwidth}m{0.98\textwidth}}
    \rowcolor[HTML]{cfe2f3}
         \multicolumn{2}{c}{
         \href{https://eur-lex.europa.eu/eli/reg/2024/1689/oj/eng\#art_11}{\large Art. 11: Technical Documentation + Annex IV}} \\ 
         & \\
         \cellcolor[HTML]{cfe2f3} \rotatebox{90}{\textbf{Interpretation guidelines}} & 
        \href{https://eur-lex.europa.eu/eli/reg/2024/1689/oj/eng#rct_71}{Recital 71}: 
        \textit{Having comprehensible information on how high-risk AI systems have been developed and how they perform throughout their lifetime is essential to enable traceability of those systems, verify compliance with the requirements under this Regulation, as well as monitoring of their operations and post market monitoring. This requires keeping records and the availability of technical documentation, containing information which is necessary to assess the compliance of the AI system with the relevant requirements and facilitate post market monitoring. Such information should include the general characteristics, capabilities and limitations of the system, algorithms, data, training, testing and validation processes used as well as documentation on the relevant risk-management system and drawn in a clear and comprehensive form. The technical documentation should be kept up to date, appropriately throughout the lifetime of the AI system. Furthermore, high-risk AI systems should technically allow for the automatic recording of events, by means of logs, over the duration of the lifetime of the system.}
         \\ 
    \end{tabular}
\end{table}

\begin{longtable}{p{0.4\textwidth}p{0.6\textwidth}}
\rowcolor[HTML]{cfe2f3}\textbf{Compliance checklist} & \textbf{Guiding questions} \\
& \\
\endhead
\multirow{5}{0.4\textwidth}{\parbox{0.4\textwidth}{11.1 The technical documentation for the high-risk AI system was prepared before the system was \textbf{placed on the market / put into service, is maintained up-to-date}, it includes at least all elements required by Annex IV, and is presented in a \textbf{clear and comprehensive form} enabling assessment of compliance with Section 2 requirements.}} &11.1.1  Did you prepare the technical documentation before placing the system on the market or putting it into service, in a way that it is clear and comprehensive and includes all elements in Annex IV? \\
&11.1.2  Did you identified any Annex IV items or Section 2 areas that could not be fully completed? \\
&11.1.3  How frequently is the technical documentation reviewed and updated? \\
&11.1.4  Do you update the technical documentation whenever there are architectural changes or integration of third-party components? \\
&11.1.5  Was there an internal assessment on the technical documentation comprehensiveness (e.g. cross-departments review)? \\
\midrule
\multicolumn{2}{l}{
\cellcolor[HTML]{efefef}
\parbox{\linewidth}{
\textit{
11.2 [Art.11.2 does not apply as the system in analysis is not referred in Annex I, Section A but in Annex III.4(a)]
}}
} \\
\midrule
\multirow{3}{0.4\textwidth}{\parbox{0.4\textwidth}{11.3 The provider \textbf{follows the delegated acts} of the Commission to amend Annex IV and promptly update the technical documentation accordingly.}} &11.3.1  Has the Commission adopted any delegated acts in accordance with Article 97 in order to amend Annex IV? \\
&11.3.2  What specific changes to Annex IV are introduced by the delegated act(s) and which parts of your technical documentation are affected? \\
&11.3.3    Does the updated technical documentation still provide all the information necessary to assess the compliance with all requirements set out Section 2? \\
\midrule\multicolumn{2}{l}{
\cellcolor[HTML]{efefef}
\parbox{\linewidth}{
\textit{
Annex IV.1. The technical documentation includes a general description of:
}}
} \\
Annex IV.1(a). The \textbf{intended purpose}, the name of the provider and the version of the AI system reflecting its relation to previous versions. &Annex IV.1.a.1 Write the intended purpose, the name of the provider and the version of the AI systems reflecting its relation to previous versions. \\
\midrule
\multirow{2}{0.4\textwidth}{\parbox{0.4\textwidth}{Annex IV.1(b). How the AI system \textbf{interacts with, or can be used to interact with, hardware or software, including with other AI systems}, that are not part of the AI system itself, where applicable.}} &Annex IV.1.b.1   Write how the AI system interacts with (or can be used to interact with) components that are not part of the system itself. Such components might be hardware or software, including other AI systems. \\
&Annex IV.1.b.2 Could such interactions affect system functionality or compliance with Section 2 requirements? If so, how? How do you address possible non-compliance? \\
\midrule
Annex IV.1(c). The versions of relevant software or firmware, and any requirements related to \textbf{version updates}. &Annex IV.1.c.1   Write the versions of relevant software and firmware used by the AI system and its requirements and procedures related to version updates. \\
\midrule
Annex IV.1(d). All the \textbf{forms} in which the AI system is placed on the market or put into service, such as software packages embedded into hardware, downloads, or APIs. &Annex IV.1.d.1   Write all the forms in which the AI system is placed on the market or put into service such as software packages embedded into hardware, downloads, or APIs. \\
\midrule
Annex IV.1(e). The \textbf{hardware} on which the AI system is intended to run. &Annex IV.1.e.1 Describe the hardware on which the AI system is intended to run. \\
\midrule
Annex IV.1(f). Where the AI system is a component of products, photographs or illustrations showing external features, the marking and internal layout of those products. &Annex IV.1.f.1     Provide photographs/illustrations showing external markings, features and internal layout of the products the AI system is embedded in. \\
\midrule
\multirow{3}{0.4\textwidth}{\parbox{0.4\textwidth}{Annex IV.1(g). A basic description of the \textbf{user-interface} provided to the deployer.}} &Annex IV.1.g.1  Describe the UI provided to the deployer. Consider providing sample screenshots, walkthroughs and UI examples mapped to representative deployer use-cases. Consider describing the UI elements that raise awareness about non visualized informations by-default (e.g., pruned list of candidates). \\
&Annex IV.1.g.2   Does your description include technical details of the UI (e.g. development language details, integration of user feedbacks through input forms)? \\
&Annex IV.1.g.3   Does your description cover limitations of your AI system, so that deployer can address anomalies, dysfunctions and unexpected anomalies? \\
\midrule
\multirow{3}{0.4\textwidth}{\parbox{0.4\textwidth}{Annex IV.1(h). \textbf{Instructions for use} for the deployer, and a basic description of the user-interface provided to the deployer, where applicable. [Also refer to Article 13.2-3]}} &Annex IV.1.h.1  Provide the instructions for use of the high-risk AI system for the deployer. \\
&Annex IV.1.h.2  Write a basic description of the user-interface provided to the deployer. [Also refer to Annex IV.1.g] \\
&Annex IV.1.h.3  Does your instruction description consider the level of technical proficiency of the deployer? \\
\midrule
\multicolumn{2}{l}{
\cellcolor[HTML]{efefef}
\parbox{\linewidth}{
\textit{
Annex IV.2. The technical documentation includes a detailed description of the elements of the AI system and of the process for its development, including:
}}
} \\
\multirow{7}{0.4\textwidth}{\parbox{0.4\textwidth}{Annex IV.2(a). The methods and steps performed for the development of the AI system, including, where relevant, \textbf{recourse to pre-trained systems or tools provided by third parties} and how those were used, integrated or modified by the provider.}} &Annex IV.2.a.1   What are the sequential development phases for this AI system (i.e., requirements gathering, data collection, annotation, prototyping, model selection, fine-tuning, evaluation), and which concrete artifacts (documents, datasets, model checkpoints, scripts, test reports) were produced at each phase? \\
&Annex IV.2.a.2 Which pre-trained systems or third-party tools, including General Purpose AI models, were used (vendor, model name/version), and how were they integrated, configured, or modified for your system? \\
&Annex IV.2.a.3 Has the third-party company that provided the AI model signed the Code of Conduct for General Purpose AI models? \\
&Annex IV.2.a.4  In absence of training and fine-tuning, what steps did you take to adapt a pre-trained model to your specific recruitment context at inference time (prompt design, input data format)? \\
&Annex IV.2.a.5   In case you use in-context learning, did you choose a fixed protocol (e.g. fixed prompts) for each candidate?  Are those prompts documented and versioned? Consider showing the prompts or template used. \\
&Annex IV.2.a.6     For any configurable inference parameters of pre-trained models (i.e., temperature, top-k, top-p, beam size), which parameter values were chosen for deployment, and why were they chosen? \\
&Annex IV.2.a.7    How do you track and enforce version coherence of third-party components in time across development? \\
\midrule
\multirow{15}{0.4\textwidth}{\parbox{0.4\textwidth}{Annex IV.2(b). The \textbf{design specifications} of the system, namely the \textbf{general logic} of the AI system and of the algorithms; the \textbf{key design choices} including the rationale and assumptions made, including with \textbf{regard to persons or groups of persons} in respect of who, the \textbf{system is intended to be used}; the main classification choices; what the \textbf{system is designed to optimise} for, and the relevance of the different parameters; \textbf{the description of the expected output and output quality of the system}; the decisions about \textbf{any possible trade-off} made regarding the technical solutions adopted to comply with the \textbf{requirements set out in Chapter III, Section 2}.}} &Annex IV.2.b.1  Write the general logic of the AI system and of the algorithms employed, as well as the assumptions about the context of use that guided the design of the AI system. \\
&Annex IV.2.b.2  Which is the specific task you expect the AI system to perform? What is its function? \\
&Annex IV.2.b.3  Which persons or groups of persons will interact with or be affected by the system, and what types of data will be collected and processed for each group? \\
&Annex IV.2.b.4  Which persons or groups of persons (demographic, linguistic, disability, cultural, or other subgroups) is the system not trained to recognise or handle appropriately, and what evidence supports this limitation? \\
&Annex IV.2.b.5  What is the target objective of the system (i.e., ranking score, accuracy), how is it computed, and why was it chosen? Does its computation address possible violations of fundamental rights? 
Do you have visibility on the whole computation process, including the substasks performed by third-party components? \\
&Annex IV.2.b.6 Which business KPIs does the system (including third-party components) optimise for (e.g., time-to-hire, cost, interview conversion), how did you verify these KPIs do not violate fundamental rights (e.g., non-discrimination, equal-opportunity)? \\
&Annex IV.2.b.7  What learning paradigms apply to (each component of) the AI system (e.g., supervised classification/regression, ranking, unsupervised clustering, generative inference)? If multiple paradigms or component exist, list them all. \\
&Annex IV.2.b.8   Which data preparation steps were taken for all variables (target and features)? For each variable, what type/format was used (e.g. continuous, categorical), how were continuous variables discretised or bucketed, and why were those decisions taken? \\
&Annex IV.2.b.9  How was output quality evaluated (which metrics, datasets, validation procedures), why were those metrics chosen, what acceptance thresholds were set, and what is the remediation procedure when performance falls below thresholds? \\
&Annex IV.2.b.10 Does the system integrates GPAI model(s)? \\
&-     If so, at which stages of the recruitment pipeline are they deployed, and for which specific functions? \\
&-     Were comparative experiments conducted to assess alternative approaches? \\
&-     Which datasets and evaluation metrics were used in these experiments, and what results supported the decision to adopt the GPAI-based solution? \\
&Annex IV.2.b.11 Why is the overall chosen technical solution the best justified option for the intended recruitment purpose and for complying with Chapter III, Section 2 requirements? \\
&Annex IV.2.b.12  Which experimental comparisons between technical solutions were performed (e.g., w.r.t. expected functioning, expected performance, impact on FRIA) and what are the key results? Which alternatives were rejected and why? \\
&Annex IV.2.b.13 In case your AI system relies on third parties APIs, meaning that the internals of such models cannot be accessed or modified, then how do you ensure that changes in the APIs do not lead to inconsistent results downstream? \\
&-     If you allow for inconsistent results in a large number of trials, are they negligible? How do you define the threshold for specifying it so? \\
&Annex IV.2.b.14 Did you assess the residual risk of the AI system? Is this explicitly communicated to deployers and affected persons (e.g., through the information for use and/or UI design)? \\
&Annex IV.2.b.15 Which trade-offs between Chapter III, Section 2 requirements were identified (e.g., accuracy vs explainability, reduction of cost through automation vs increased control through human oversight)? What implementation choices did you make to balance them, and what (if anything) was intentionally left out? Is this explicitly communicated to deployers and affected persons (e.g., through the information for use and/or UI design)? \\
\midrule
\multirow{6}{0.4\textwidth}{\parbox{0.4\textwidth}{Annex IV.2(c). The description of the \textbf{system architecture} explaining how software components build on or feed into each other and integrate into the overall processing; the \textbf{computational resources} used to develop, train, test and validate the AI system.}} &Annex IV.2.c.1    Describe the system architecture, its components and how they feed into each other and integrate into the overall processing. \\
&Annex IV.2.c.2     Which in-house or third-party hardware and/or cloud resources (CPUs, GPUs, memory, storage) supported each stage, and how do you ensure resource availability and reproducibility? \\
&Annex IV.2.c.3     Do the hardware and/or cloud resources come from external vendors? If so, do they provide reports on the resource usage? \\
&Annex IV.2.c.4     Can you indicate the computational resources used to develop, test and validate the AI system? Please also include computational resources for additional inference (e.g. long prompts, in-context learning, RAG). \\
&Annex IV.2.c.5     How do you isolate environments (development, testing, production) and prevent configuration drift? \\
&Annex IV.2.c.6  If you consider environmental sustainability in the development, testing and deployment of your system, list the computational resources of the different tested solutions with respect to the chosen one. \\
\midrule
\multicolumn{2}{l}{
\cellcolor[HTML]{efefef}
\parbox{\linewidth}{
\textit{
Annex IV.2(d). Where relevant, the data requirements in terms of datasheets describing the training methodologies and techniques and the training data sets used, including a general description of these data sets, information about their provenance, scope and main characteristics; how the data was obtained and selected; labelling procedures (e.g. for supervised learning), data cleaning methodologies (e.g. outliers detection). [Not applicable to our specific case, given the absence of training]
}}
} \\
\midrule
\multirow{6}{0.4\textwidth}{\parbox{0.4\textwidth}{Annex IV.2(e). Assessment of the \textbf{human oversight measures} needed in accordance with Article 14, including an assessment of the \textbf{technical measures needed to facilitate the interpretation of the outputs} of AI systems by the deployers, in accordance with \textbf{Article 13(3), point (d)}.}} &Annex IV.2.e.1     Describe and justify the assessment of all the human oversight measures required in Article 14. \\
&Annex IV.2.e.2 Describe which Human-Machine Interaction elements (pause/stop/override/defer, feedback mechanisms, shutdown) you defined and how are they implemented. \\
&Annex IV.2.e.3 When emergent risks to health, safety, or fundamental rights occur, which operational parameters (such as latency budgets or workload/queue constraints) are required to guarantee timely human intervention? [For completeness, refer also to guiding questions of Article14.] \\
&Annex IV.2.e.4     Describe and justify the assessment of all the technical measures needed to facilitate the interpretation of the output by the deployers. \\
&- Describe the output of the system, including confidence/uncertainty, limitations, rationale/explanation, data-quality flags if present. \\
&- Describe the implemented interpretability tools (e.g., calibrated scores, example-based aids, explanations) and their limitations. \\
&Annex IV.2.e.5     How did you assess whether the deployers were able to properly interpret the output of the AI system? \\
&- What tests/drills/usability studies demonstrate that overseers can detect anomalies and override/stop the system? \\
&- What measures against automation bias/over-reliance are in place, and what evidence shows they work (e.g., A/B or usability results)? \\
&Annex IV.2.e.6  When the model/data/UI changes, how are oversight and interpretability re-assessed and re-documented? \\
\midrule
\multirow{4}{0.4\textwidth}{\parbox{0.4\textwidth}{Annex IV.2(f). Where applicable, a detailed description of \textbf{pre-determined changes} to the AI system and its performance, together with all the relevant information related to the \textbf{technical solutions adopted to ensure continuous compliance of the AI system} with the relevant requirements set out in Chapter III, Section 2.}} &Annex IV.2.f.1     Describe the foreseeable pre-determined changes to the AI system. \\
&-  Have you classified types of predetermined changes (data, model, thresholds, feature pipeline, UI/HMI, integrations/APIs, third-party components, security patches) and mapped each to documentation locations? \\
&-  What is the boundary between a pre-determined change and a ``substantial modification'' (i.e., Article 3(23)) that forces new conformity steps? State criteria and decision path for each identified foreseeable pre-determined change. \\
&Annex IV.2.f.2     If present, describe the self-learning mechanisms (i.e., incremental model updates, threshold adjustments) that were approved during initial conformity assessment. \\
&-  For self-learning updates (online/incremental), are learning modes bounded (on/off by default, rate limits, whitelist features), and how are drift triggers handled? \\
&-  Are there tests performed to prove that the adjustment due to self-learning mechanisms do not impact the initial conformity assessment? \\
&-   For each self-learning mechanism, what are the guardrails: data sources allowed, update cadence, max parameter drift, performance floors (i.e., accuracy/robustness/fairness/security), and compatibility constraints (e.g., between software and hardware components)? \\
&-   What are the automatic rollback and safe-fail/hold conditions if guardrails are breached? \\
&Annex IV.2.f.3     In case of pre-determined changes to the system, how is compliance re-validated? \\
&- Which type of changes are automatically implemented and which ones require human sign-off? \\
&Annex IV.2.f.4 After compliance re-assessment, how did you update the technical documentation, post market monitoring plan and/or instructions for use? \\
&- If pre-determined changes occurred compared to a previous version of the system, document any tradeoff in alignment with the requirements set out in Chapter III, Section 2 (e.g. reduction in accuracy, robustness or cybersecurity, worsening of discriminatory impacts) providing before/after evidence. \\
\midrule
\multirow{8}{0.4\textwidth}{\parbox{0.4\textwidth}{Annex IV.2(g). The \textbf{validation and testing procedures} used, including information about the validation and testing data used and their main characteristics; \textbf{metrics used to measure} accuracy, robustness and compliance with other relevant requirements set out in Chapter III, Section 2, as well as \textbf{potentially discriminatory impacts; test logs and all test reports} dated and signed by the \textbf{responsible persons}, including with regard to predetermined changes as referred to under point (f).}} &Annex IV.2.g.1     Describe and justify the validation and testing procedures used. \\
&Annex IV.2.g.2     Describe and justify the validation and testing data used. \\
&- Describe the target variable in validation and testing data. \\
&- Describe the data annotation and labelling procedures, including interactive techniques and crowdsourcing tasks if used. \\
&Annex IV.2.g.3 What are the shared criteria given to annotators for data labelling (e.g., which are the guidelines for assessing ``best'' candidate) and how is this affected by their subjectivity? \\
&Annex IV.2.g.4 Has data labelling performed by domain experts (i.e. recruiters)? \\
&Annex IV.2.g.5     Do you find that your labelling procedure falls outside of the scope of objective labelling? (i.e., who defines the score/ ground truth? Is it a subjective ground truth? Do you assess the objectivity of such ground truth? Is the final user of the AI system the original labeler?) \\
&Annex IV.2.g.6     Which metrics did you use to quantitatively evaluate at test time accuracy, robustness and compliance with other relevant requirements set out in Chapter III, Section 2? Are such metrics able to fully capture both potential and limitations of the AI systems? \\
&Annex IV.2.g.7     Include all test logs and reports regarding validation and testing procedures dated and signed by the responsible persons. \\
&Annex IV.2.g.8     Include all test logs and test reports regarding predetermined changes to the AI system dated and signed by the responsible persons. \\
\midrule
\multirow{3}{0.4\textwidth}{\parbox{0.4\textwidth}{Annex IV.2(h). \textbf{Cybersecurity measures} put in place.}} &Annex IV.2.h.1     Describe the cybersecurity measures did you put in place to ensure the resilience of the system. \\
&Annex IV.2.h.2     Describe the cybersecurity vulnerabilities of your system. \\
&\cellcolor[HTML]{efefef}\textit{Please Refer to Article15 (5) for detailed cybersecurity guiding questions.} \\
\midrule
\multicolumn{2}{l}{
\cellcolor[HTML]{efefef}
\parbox{\linewidth}{
\textit{
Annex IV.3-9. A detailed information and/or description about:
}}
} \\
\multirow{6}{0.4\textwidth}{\parbox{0.4\textwidth}{Annex IV.3(1°part). The \textbf{monitoring}, \textbf{functioning} and \textbf{control} of the AI system, in particular with regard to: its \textbf{capabilities and limitations in performance}, including the \textbf{degrees of accuracy for specific persons or groups of persons} on which the system is intended to be used and the overall \textbf{expected level of accuracy} in relation to its intended purpose.}} &Annex IV.3.1   Describe and justify the monitoring, functioning and control of the AI system. \\
&Annex IV.3.2     Provide a detailed description of its capabilities and limitations in performance. \\
&Annex IV.3.3     Provide a detailed description of the degree of accuracy and its continuous assessment for specific persons or groups of persons for whom the system is intended. \\
&Annex IV.3.4  In case of accuracy degradation, detail the mitigation procedure in order to recover the allowed degree of accuracy. \\
&Annex IV.3.5     Did you document the degree of accuracy reached, overall and for specific group, in each sub-task of your application (e.g. CV parsing and Candidate Scoring)? \\
&Annex IV.3.6     What are the minimum confidence thresholds below which results are deemed unreliable in the real intended context? And why? \\
\midrule
\multirow{3}{0.4\textwidth}{\parbox{0.4\textwidth}{Annex IV.3(2°part). The monitoring, functioning and control of the AI system, in particular with regard to: the \textbf{foreseeable unintended outcomes} and \textbf{sources of risks to health and safety, fundamental rights} and discrimination in view of the intended purpose of the AI system.}} &Annex IV.3.7    How will you monitor and control the sources of risks to health, safety and fundamental rights in view of the intended purpose of the AI system? \\
&Annex IV.3.8     Which unintended outcomes, environmental or contextual factors have you identified that could degrade performance and/or compromise safety? \\
&Annex IV.3.9     How will you monitor and control the foreseeable unintended outcomes in view of the intended purpose of the AI system? \\
\midrule
\multirow{4}{0.4\textwidth}{\parbox{0.4\textwidth}{Annex IV.3(3°part). The monitoring, functioning and control of the AI system, in particular with regard to: the \textbf{human oversight measures} needed in accordance with Article 14, including the \textbf{technical measures put in place to facilitate the interpretation of the outputs of AI systems} by the \textbf{deployers}.}} &Annex IV.3.10     Did you describe all human oversight measures needed in accordance with Article 14? [For completeness, refer also to guiding questions of Article 14.] \\
&Annex IV.3.11     Did you describe and justify all the technical measures needed to facilitate the interpretation of the output by the deployers? \\
&Annex IV.3.12     For instance, did you consider any explainability techniques that can facilitate the interpretation of the output? [For completeness, refer also to guiding questions on explainability in Article 13(1).] \\
&Annex IV.3.13     Did you consider any other interpretable-by-design models that can facilitate the interpretation of the output? \\
\midrule
\multirow{5}{0.4\textwidth}{\parbox{0.4\textwidth}{Annex IV.3(4°part). The monitoring, functioning and control of the AI system, in particular with regard to: \textbf{specifications on input data}, as appropriate.}} &Annex IV.3.14     Describe the specifications on the input data that you expect the AI system to be fed with. \\
&Annex IV.3.15     Have you verified that the input data used in practice fall within the boundaries (formats, ranges, sources, statistical properties) defined during validation and testing, and that they are sufficiently representative to ensure that the expected performance and risk profile remain valid? \\
&Annex IV.3.16     Given your design, do you anticipate any deviations in the input data relative to the validation and testing data? How are these deviations detected and addressed? \\
&Annex IV.3.17  Do you expect conditions under which the input data may present characteristics such that the AI system will treat them as anomalies? \\
&Annex IV.3.18     Which data are expressively prohibited (e.g. in terms of format and/or content), and how are violations detected and logged? \\
\midrule
\multirow{8}{0.4\textwidth}{\parbox{0.4\textwidth}{Annex IV.4. \textbf{Appropriateness} of the \textbf{performance metrics} for the specific AI system.}} &Annex IV.4.1   Describe and justify the choice of the performance metrics used to evaluate and monitor the AI system. \\
&-     Are they appropriate for the intended purpose of the system? \\
&-     What are their limitations? \\
&-     How did you define acceptance thresholds for each metric? \\
&-     How do you define acceptable risk levels for each metric? \\
&-     How do the chosen metrics relate to the risks identified under Article 9 (i.e., risk management system)? \\
&-     Do the metrics capture performance across different persons or groups of persons on which the system is intended to be used? \\
&-     How will the metrics be monitored and updated post-market to detect drift or emerging risks? \\
\midrule
\multirow{1}{0.4\textwidth}{\parbox{0.4\textwidth}{Annex IV.5 \textbf{Risk management system} in accordance with Article 9.}} &Annex IV.5.1     Describe and justify the risk management system in accordance with Article 9. [For completeness, refer also to guiding questions of Article 9] \\
\midrule
\pagebreak
\multirow{3}{0.4\textwidth}{\parbox{0.4\textwidth}{Annex IV.6. \textbf{Relevant changes} made by the provider to the system through its \textbf{lifecycle}.}} &Annex IV.6.1     How do you identify relevant changes? \\
&-     What is your definition of relevant change, w.r.t the context of application? \\
&-     How do you distinguish relevant changes from minor changes? \\
&-     For each change what do you record and log? (i.e., date, responsible person, moment in the lifecycle, justification) \\
&Annex IV.6.2     As a provider of the AI system, did you make any relevant changes to the AI system? \\
&-     If so, describe and justify them. \\
&-     How do these changes impact the compliance with the requirements set out in Chapter III, Section 2? \\
&-     Do such relevant changes qualify as substantial modifications? \\
\midrule
\multirow{2}{0.4\textwidth}{\parbox{0.4\textwidth}{Annex IV.7. \textbf{Harmonised standards} published in the Official Journal of the European Union apply in full or in part to your high risk AI system. If no such harmonised standards have been applied, a detailed description of the solutions adopted to meet the requirements of Chapter III, Section 2, including a list of other relevant standards and technical specifications applied has been provided.}} &Annex IV.7.1     Which harmonised standards published in the Official Journal of the European Union apply in full or in part to your high-risk AI system? \\
&Annex IV.7.2     If no such harmonised standards have been applied, or applied only partially, have you provided a detailed description of the solutions adopted to meet the requirements of Chapter III, Section 2? \\
& \\
& \\
& \\
\midrule
Annex IV.8. A copy of the \textbf{EU Declaration of Conformity} referred to in Art. 47. &Annex IV.8.1     Include a copy of the EU Declaration of Conformity (Article 47) \\
\midrule
\multirow{5}{0.4\textwidth}{\parbox{0.4\textwidth}{Annex IV.9. System in place to evaluate the AI system performance in the \textbf{post-market phase} in accordance with Article 72, including the post-market monitoring plan referred to in Article 72(3).}} &Annex IV.9.1     How did you evaluate the AI system performance in the post-market phase (Article 72)? [For completeness, refer also to guiding questions on record-keeping of Article12] \\
&Annex IV.9.2     What is your market monitoring plan (Article 72(3))? Describe it and justify the choice of the specific plan (e.g., in terms of attention points, frequency, metrics employed, human involvement, ect). \\
&Annex IV.9.3 Describe how the monitoring plan detects data drift, performance degradation, or emerging risks to health, safety, or fundamental rights. \\
&Annex IV.9.4 What mechanisms are in place to ensure that monitoring results fed back into the risk management system (Article 9) and that corrective actions are triggered where necessary? \\
&Annex IV.9.5 How do you integrate feedback from deployers, users, or incident reports (Article 73) into your post-market evaluation? \\
\bottomrule 

\end{longtable}
    \par\mbox{}\clearpage
    \begin{table}[h]
    \centering
    \begin{tabular}{m{0.02\textwidth}m{0.98\textwidth}}
    \rowcolor[HTML]{cfe2f3}
         \multicolumn{2}{c}{
         \href{https://eur-lex.europa.eu/eli/reg/2024/1689/oj/eng\#art_12}{\large Art. 12: Record-keeping}} \\ 
         & \\
         \cellcolor[HTML]{cfe2f3} \rotatebox{90}{\textbf{Interpretation guidelines}} & 
         \href{https://eur-lex.europa.eu/eli/reg/2024/1689/oj/eng#rct_71}{Recital 71}: 
        \textit{Having comprehensible information on how high-risk AI systems have been developed and how they perform throughout their lifetime is essential to enable traceability of those systems, verify compliance with the requirements under this Regulation, as well as monitoring of their operations and post market monitoring. This requires keeping records and the availability of technical documentation, containing information which is necessary to assess the compliance of the AI system with the relevant requirements and facilitate post market monitoring. Such information should include the general characteristics, capabilities and limitations of the system, algorithms, data, training, testing and validation processes used as well as documentation on the relevant risk-management system and drawn in a clear and comprehensive form. The technical documentation should be kept up to date, appropriately throughout the lifetime of the AI system. Furthermore, high-risk AI systems should technically allow for the automatic recording of events, by means of logs, over the duration of the lifetime of the system.}
         \\ 
    \end{tabular}
\end{table}

\begin{longtable}{p{0.4\textwidth}p{0.6\textwidth}}
\rowcolor[HTML]{cfe2f3}\textbf{Compliance checklist} & \textbf{Guiding questions} \\
& \\
\endhead
\multirow{17}{0.4\textwidth}{\parbox{0.4\textwidth}{1. The high-risk AI system \textbf{technically allows} for the automatic recording of events (logs) over its own \textbf{lifetime}.}} &12.1.1     Which system events are captured automatically (e.g., API calls, model inferences, error conditions)? \\
&-     Are there any system events that are not captured automatically? If so, why? \\
&12.1.2     What is the level of granularity decided for the system logging? (i.e., metadata only or full payload)? \\
&12.1.3     Has a logging standardized format been defined to ensure consistency? \\
&12.1.4     Is the logging of contextual metadata (timestamps, model version, pipeline stage) enough to allow for a full reconstruction of each event and possible chain of events as required by Art.20? \\
&12.1.5     Where the AI system relies on external AI or GPAI components during its operation, are logging capabilities designed to record the invocation and role of those components in system outputs (e.g., external inference requests, returned outputs, component identifiers or version information), so that their contribution can be reconstructed in case of risk investigation or system modification?\\
&12.1.6     How will the logs be stored (e.g. on-premise, cloud storage)? \\
&12.1.7     Is there a detection system to prevent unauthorized modification or deletion of log entries according to the storage environment? Describe it and justify its efficacy. \\
&12.1.8     Is there a detection system to alert on any attempt to tamper with the logging system? \\
&-     Did you define log patterns or anomalies that should trigger automated alerts? \\
&-     How will alert notifications be routed (i.e., to on-call engineers, compliance officers) and tracked to resolution? \\
&12.1.9     How are the durability and scalability of the logging storage addressed during the design phase and management over time? \\
&12.1.10     Are access rights and roles defined to specify who (users and/or systems) can read, query, or export logs? \\
&12.1.11     Is there an expected performance impact of logging? \\
&-     If yes, which strategies will you employ to minimize overhead without losing critical system events? \\
&12.1.12     Are the logs indexed in order to allow for search in case of incident investigation, audit or debugging? \\
&12.1.13     Do you integrate any real-time monitoring system for the event logs? \\
&-     If yes, does it consider meta-data or the log content itself? Justify your choice, ensuring there is no significant loss of information that could impact the efficacy of the monitoring. \\
&12.1.14     Do you integrate tools or dashboards to analyze historical data of the system logs? \\
&-     If yes, does it consider meta-data or the log content itself? Justify your choice, ensuring there is no significant loss of information that could impact the efficacy of the monitoring. \\
&12.1.15 Have you established an internal or third-party audit process for the logging system to verify completeness and compliance? \\
&-          If yes, have you also defined the time frequency of such audits? \\
&12.1.16     Do you generate reports to demonstrate to regulators that logs are collected and stored correctly (e.g. protected in terms of cybersecurity, compliant with GDPR)? If yes, how are such reports generated? \\
&12.1.17     Have you defined a process that governs possible updates of the log schema as new features (or regulations) emerge? \\
&-     If yes, how do you ensure backward compatibility so that past logs remain consistent with the new schema? \\
\midrule\multirow{32}{0.4\textwidth}{\parbox{0.4\textwidth}{2(a). The high-risk system records events (logs) \textbf{relevant to identify} situations in which the system may present a risk (Art. 79(1)) or has undergone a substantial modification (Art. 3(23)), with the goal of \textbf{ensuring a level of traceability} of the functioning of a high-risk AI system that is \textbf{appropriate} to its intended purpose.}} &\cellcolor[HTML]{efefef}\textit{— Art.79(1) —} \\
&12.2.a.1     What types of system outputs or patterns do you treat as indicators that the system might be presenting a risk (e.g., unusually high false-negative rates in candidate screening, repeated demographic bias flags, error rates above a predefined threshold, or significant deviations from expected values)? \\
&\cellcolor[HTML]{efefef}\textit{[The above question presumes that each point of the example list is defined and/or can be measured and/or can be tracked in a meaningful way; examples are the following subquestions:]} \\
&-     What are the numerical thresholds for a ``risk-presenting'' event triggering a log entry (e.g., confidence score $<$ x, false-positive rate $>$ y)? \\
&-    Do you log every instance of a borderline decision, and if so, which specific cases are included (e.g., a candidate ranked just below a cutoff)? Or do you instead log only aggregate anomaly spikes over time (e.g., recurring spikes in error rates)? \\
&12.2.a.2     When defining logging triggers, how do you capture the distinction between the system's intended purpose (Art. 3(12)) and reasonably foreseeable misuse (Art. 3(13))? For example, if a recruitment AI system is applied outside its validated demographic scope, is this treated and logged as a misuse event? \\
&-     Which specific misuse scenarios have you identified for your system, and how are these scenarios monitored and logged? \\
&12.2.a.3     Is the granularity of logs defined in such a way that it is possible to reconstruct whether a given decision was within the system's intended use (e.g., logging that the candidate's origin was outside the originally approved demographic)? \\
&12.2.a.4     For each flagged risk event, which contextual attributes do you record to allow a retrospective ``root-cause'' analysis (e.g., input snapshot, model version, pipeline steps, timestamp, user or service ID that invoked the decision, or the LLM in-context learned snapshot up to the risk event)? \\
&12.2.a.5     How do you log instances where the system is used in ways that deviate from its normal or expected use? (For example, if a recruitment AI system designed for screening engineer-level candidates is applied to executive-level profiles.) \\
&12.2.a.6     Once a risk-presenting event is logged, who must be notified? \\
&12.2.a.7     Did you define a priority hierarchy among the notified persons (e.g., first the provider, then the recruiters)? \\
&12.2.a.8     Which criteria trigger an immediate real-time alert versus a periodic report? \\
&12.2.a.9     How are alerts routed (e.g., email, ticketing, SMS to on-call engineer)? \\
&12.2.a.10     What downstream workflows must be invoked once a risk log is created (e.g., automatic freeze of new decisions pending human review)? \\
&\cellcolor[HTML]{efefef}\textit{— Art.3(23) —} \\
&12.2.a.11     Which system changes do you classify as non-substantial modifications, meaning they do not alter the intended purpose or affect compliance with Chapter III Section 2 requirements? (e.g., accuracy and fairness assessment metrics remain within the bounds approved during initial conformity assessment) \\
&12.2.a.12     Which system modifications would you classify as falling outside the scope of pre-approved behaviours, and therefore potentially count as substantial modifications (e.g., replacing the underlying LLM model architecture with a new version)? \\
&12.2.a.13     How do you distinguish ``configuration modifications'' (e.g., hyperparameter tuning with pre-approved ranges) from ``substantial modifications''? What metadata do you record to prove that a modification was within allowable bounds? \\
&12.2.a.14     Do you have rules in place to detect at run-time when updates or changes to system components (e.g., pipeline packages, libraries, or other dependencies) qualify as substantial modifications? \\
&12.2.a.15     For a given substantial modification, which information is logged to ensure traceability? A non-exhaustive list should include: ``modification'' related information (e.g., switch to new inference engine); ``versioning'' related information (e.g., previous model -> new model version, pipeline code commit, if applicable pipeline container images); timestamps of activation of the modification; reason and/or justification for the modification (e.g., performance drift detected, feature engineering update); operator/actor who initiated the modification; affected components of the change (pipeline components, decision logics). \\
&12.2.a.16    After a modification, which checks or automated tests must pass to demonstrate that the system continues to meet Chapter III, Section 2 requirements? (For example, bias audits, robustness tests, or accuracy benchmarks.) \\
&12.2.a.17     How do logs capture the results of such checks? (e.g., pass/fail, details, test metrics) \\
&12.2.a.18     Once a modification is logged as substantial, is there an automated process that ensures that the AI product is flagged for a new conformity assessment? \\
&-          Is such a log entry linked to an internal/external audit ticket so that the compliance team can demonstrate traceability from the ``modification event'' to ``audit outcome''? (e.g. AI version $\rightarrow$ substantial change log $\rightarrow$ new conformity assessment documents?) \\
&12.2.a.19 Once a modification event is logged, who is notified and who is authorized to see it? In case the modification is deemed as substantial, who is notified? \\
&12.2.a.20     How do you ensure that logs of risk-presenting events are fed back into system development and design improvements? For instance, if a specific model version consistently triggers risk flags, how are these logged insights transferred to and acted upon by the development team? \\
&-         Which governance process ensures that logged risks systematically inform future design choices? \\
&12.2.a.21     If you willingly perform a substantial modification, how do you ensure that this event is logged? Which information do you log? For instance: \\
&-     How do you record who initiated and approved the modification? \\
&-     How do you capture the exact time and date of the modification? \\
&-    How do you retain both the state of the system before the modification and the state after it? \\
&-    How do you document which components of the system were altered and in what way? \\
&-   How do you guarantee that these logs are preserved in a tamper-evident and auditable format? \\
&12.2.a.22     If the substantial modifications refer to a pre-allowed range of hyperparameters how can you trace those values in real-time using the logs? \\
\midrule\multirow{11}{0.4\textwidth}{\parbox{0.4\textwidth}{2(b). The high-risk system records events (logs) \textbf{relevant} for facilitating the \textbf{post-market monitoring} referred to in Art. 72.}} &12.2.b.1     Which performance metrics (e.g., accuracy, precision, recall, demographic parity gaps) do you track to demonstrate ongoing compliance with Chapter III, Section 2? \\
&12.2.b.2     How do you log the indicators that are specific to each regulatory requirement (bias mitigation, robustness, transparency)? \\
&12.2.b.3     Does your logging system track any documents or relevant data that deployers (e.g., HR teams) might supply to feed post-market monitoring (e.g., candidate appeals, adverse outcomes reported)? \\
&12.2.b.4     Have you codified a ``post-market monitoring plan'' for logs that explicitly includes: i) Data sources (deployer feedback, in-product metrics), ii) collection frequency (e.g., real-time vs. daily batch), and iii) analysis pipeline (e.g., automated drift detection scripts)? \\
&-     For each element of the post-market monitoring plan, how do you ensure that its execution is reflected in the logs? For example, if the plan includes a daily score-alignment audit, do the logs show that the audit was performed and what its results were? \\
&12.2.b.5     How do you record the eventual updates of the post-market monitoring plan? \\
&12.2.b.6    What metadata is logged to show that at any point in time, a specific version of the post-market plan was in effect? \\
&12.2.b.7     Who (role/title) is responsible for maintaining the post-market monitoring plan? \\
&12.2.b.8     How is accountability divided between the provider and the deployer (HR, IT) for collecting and analyzing the logs related to post-market monitoring? \\
&12.2.b.9     How do insights from log analyses (e.g., emerging bias patterns in specific job categories) feed back into updates of the post-market monitoring plan itself? \\
&12.2.b.10    What technical measures in your logging system ensure that revisions of the post-market monitoring plan are recorded in a way that is protected from unauthorized change, so that the timing and content of each revision can be reliably verified during an audit? \\
&12.2.b.11     Is your post-market monitoring system integrated with existing logging frameworks? If so, how do you ensure that the fields in your logs correspond to the required template elements of those frameworks? \\
\midrule\multirow{13}{0.4\textwidth}{\parbox{0.4\textwidth}{2(c). The high-risk system records events (logs) \textbf{relevant} for \textbf{monitoring the operation} of high-risk AI systems referred to in Art.26(5). [identify risk situations, record suspensions, document notifications of serious incidents]}} &12.2.c.1     How do you verify that each operation of the system (e.g., candidate screening run, hyperparameter change) is executed according to the published Instructions for Use required by Art.13? \\
&12.2.c.2     How is a deployer's decision to suspend the AI system logged? \\
&12.2.c.3     What minimum information must be recorded to show that use was suspended ``without undue delay'' once a risk was identified (e.g., time elapsed between the first risk-flag log and the suspension log)? \\
&12.2.c.4     Which confirmation or acknowledgment is associated with suspension, and how is that captured? \\
&12.2.c.5     What log entries record the act of ``informing the provider or distributor and the relevant market surveillance authority without undue delay''? \\
&12.2.c.6    What constitutes a ``serious incident'' in your specific recruitment application context? Does it have a specific logging template and alert system? \\
&12.2.c.7     When you notify the provider and the market surveillance authority about a suspension or a serious incident, how do you log that the notification was successfully delivered and acknowledged, and how do you record any follow-up actions if the first attempt at contact fails? \\
&12.2.c.8     When the criteria for a serious incident are met, what additional fields do you log (e.g., incident ID, affected job category, number of impacted candidates, severity rating) to demonstrate compliance with Article 73 timelines? \\
&12.2.c.9     When a serious incident is detected, how do you log the point at which you first identified a possible causal link between the system and the incident, and how do you then log the reasoning and evidence that led to the conclusion that a causal link is established or likely? \\
&12.2.c.10     How do your logs capture the severity and nature of each incident, so that both immediate impact and systemic implications can be assessed later? \\
&12.2.c.11     After a serious incident log is created, which events record the subsequent ``investigation'' steps (e.g., analysis runs, interviews, and risk assessments)? \\
&12.2.c.12  During the investigation of a serious incident, how do you log that evidence (such as raw logs, datasets, or model snapshots) has been preserved without alteration before notifying authorities? \\
&12.2.c.13     When a corrective action is implemented in response to a serious incident, is it captured in the logs to ensure auditability? \\
\bottomrule 

\end{longtable}
    \par\mbox{}\clearpage
    \begin{table}[h]
\centering
\begin{tabular}{m{0.02\textwidth}m{0.98\textwidth}}
\rowcolor[HTML]{cfe2f3}
 \multicolumn{2}{c}{
 \href{https://eur-lex.europa.eu/eli/reg/2024/1689/oj/eng\#art_13}{\large Art. 13: Transparency and Provision of Information to Deployers}} \\ 
 & \\
 \cellcolor[HTML]{cfe2f3} \rotatebox{90}{\textbf{Interpretation guidelines}} & 
 \href{https://eur-lex.europa.eu/eli/reg/2024/1689/oj/eng#rct_72}{Recital 72}: 
 \textit{To address concerns related to opacity and complexity of certain AI systems and help deployers to fulfil their obligations under this Regulation, transparency should be required for high-risk AI systems before they are placed on the market or put it into service. High-risk AI systems should be designed in a manner to enable deployers to understand how the AI system works, evaluate its functionality, and comprehend its strengths and limitations. High-risk AI systems, should be accompanied by appropriate information in the form of instructions of use. Such information should include the characteristics, capabilities and limitations of performance of the AI system. [...]}
 \textit{Transparency, including the accompanying instructions for use, should assist deployers in the use of the system and support informed decision making by them. [...]}
 \\ 
\end{tabular}
\end{table}

\begin{longtable}{p{0.4\textwidth}p{0.6\textwidth}}
\rowcolor[HTML]{cfe2f3}\textbf{Compliance checklist} & \textbf{Guiding questions} \\
& \\
\endhead
\multirow{17}{0.4\textwidth}{\parbox{0.4\textwidth}{1. The high-risk system is designed and developed with sufficient transparency to enable the \textbf{appropriate use of the system and the interpretation of its output by deployers}. The type and degree of transparency must ensure compliance with the obligations of the provider and deployer given in \textbf{Section 3}.}} & 13.1.1 During the design phase of your system, did you explicitly address how deployers will interpret and act on outputs?\\
&13.1.2 Is your architecture interpretable-by-design or a blackbox?\\ 
&13.1.3 Considering the adopted architecture and intended context of use, what level of transparency does your application provide, and what are its main limitations?\\
&13.1.4 How do you define a sufficient and appropriate degree of transparency in your application?\\
&13.1.5 How did you validate your definition of “appropriate degree of transparency” with final deployers (i.e. recruiters)?\\
&13.1.6 Once you have defined the target transparency level, how do you implement it in code and/or UI?\\
&13.1.7 What is the output of your AI system? Is it represented by a single value (i.e., prediction score, ranking) or by a set of information (e.g., confidence scores, explanations)?\\
&13.1.8 Does the information provided by the AI system to ensure transparency and interpretability accurately reflect the algorithmic process that generated the output? How is the accuracy of this representation verified?\\
&13.1.9 Beyond built‑in transparency tools, what training materials, quick‑start guides, or “interpretation playbooks” do you provide to deployers?\\
&13.1.10 How do you verify that deployers correctly interpret the output of the AI system and use it appropriately within their decision-making process (e.g., post-deployment surveys)? In the case of a negative response, how do you correct the interpretability design?\\
&13.1.11 How and how often do you audit or test the interpretability of the AI system (e.g., A/B tests with varying information payload depths, simulated misuse scenarios)?\\
&13.1.12 How do you communicate to deployers (i) the residual opacity of the AI system and (ii) the limitations of the adopted interpretability techniques?\\
&13.1.13 Do you use techniques coming from Explainable AI literature to ensure an appropriate degree of transparency and interpretability of the outputs and of the process generating them?\\
&- Which techniques do you use and why?\\
&- How do you ensure that those techniques give consistent interpretations?\\
&13.1.14 Do you provide deployers with interpretability tools to correctly interpret both the job-matching score (i.e., intermediate output in the AI architecture) and the relative ranking given by scores on different individuals (i.e., final output)?\\
&13.1.15 In the case of candidates whose scores cannot be differentiated in a statistically significant manner, how is this information communicated to deployers?\\
& 13.1.16 If the AI system has been developed by integrating GPAI models, what transparency measures have been adopted, taking into account the specific characteristics of such models? What technical measures and contractual safeguards have been put in place?\\
\midrule
\multirow{4}{0.4\textwidth}{\parbox{0.4\textwidth}{2. Appropriate instructions to use the high-risk AI system are provided in an appropriate digital format or otherwise, to achieve compliance with the relevant obligations. These instructions are concise, complete, correct and clear as well as relevant, accessible and comprehensible to deployers.}}
&13.2.1 In which format are the instructions for use provided to deployers? And what is their content?\\
&- Which information is provided to interpret the output of the system in order to ensure transparency of the AI application?\\
&13.2.2 What level of AI literacy is required to understand the provided instructions for use?\\
&13.2.3 Did you receive feedback from deployers about the quality of the instructions for use (i.e., is such information concise, complete, correct, clear, relevant, accessible, and comprehensible)?\\
\midrule
\multicolumn{2}{l}{\cellcolor[HTML]{efefef}\textit{3. The \textbf{instructions for use} shall contain at least the following information:}} \\
\multirow{3}{0.4\textwidth}{\parbox{0.4\textwidth}{3(a). the identity and the contact details of the provider and, where applicable, of its authorised representative.}}
&13.3.a.1 Are the identity and contact details of the provider included in the instructions given alongside your application?\\
&13.3.a.2 If applicable in your application, are the identity and contact details of the authorised representative of the provider given in the instructions?\\
&13.3.a.3 Does the instructions for use contain the identity and the contact details of the upstream provider in the case of integration of a third-party model? If not, why?\\
\midrule
\multicolumn{2}{l}{\cellcolor[HTML]{efefef}\textit{3(b). the characteristics, capabilities and limitations of performance of the high-risk AI system, including:}}\\
3(b)(i). the intended purpose of the high-risk AI system;
&13.3.b.1 Do the instructions include the intended purpose of the system?\\
\midrule
\multirow{8}{0.4\textwidth}{\parbox{0.4\textwidth}{3(b)(ii). the level of accuracy, including its metrics, robustness and cybersecurity referred to in Article 15 against which the high-risk AI system has been tested and validated and which can be expected, and any known and foreseeable circumstances that may have an impact on that expected level of accuracy, robustness and cybersecurity;}}
&13.3.b.2 Do the instructions include the level of accuracy of your system, including the metrics, robustness and cybersecurity, against which your system has been tested and validated?\\
&13.3.b.3 Do the instructions include any known and foreseeable circumstances that may affect the level of accuracy, robustness and cybersecurity?\\
&13.3.b.4 Do your instructions state and justify any problem and limitations in defining metrics regarding accuracy, robustness and cybersecurity in the test or validation phase?\\
&13.3.b.5 Do your instructions include some guidelines on how to handle foreseeable and/or unforeseeable drops in the level of accuracy, robustness and cybersecurity?\\
&- What is the expected workload for the implementation of these guidelines? For example, low effort would indicate that every deployer can perform the operations induced by the guidelines. Justify your assumptions about the workload.\\
&- Did you previously verify with your deployer that these guidelines are actionable?\\
&13.3.b.6 Is there an update mechanism to add unforeseen circumstances in the instructions for use after they have been identified in deployment?\\
&\cellcolor[HTML]{efefef}\textit{For completeness, please refer to Art.15.}\\
\midrule
\multirow{6}{0.4\textwidth}{\parbox{0.4\textwidth}{3(b)(iii). known or foreseeable circumstances related to the use of the high-risk AI system according to the intended purpose or reasonably foreseeable misuse which may lead to health, safety, or fundamental rights risks (Article 9(2));}}
&13.3.b.7 Do the instructions include any known or foreseeable circumstances which may lead to risks to the health and safety or fundamental rights\\
&- related to the use of your AI system based on the intended purpose?\\
&- related to the reasonably foreseeable misuse of your system?\\
&13.3.b.8 For each of the known and foreseeable circumstances which may lead to risks to the health and safety or fundamental rights, do you have an explicit plan on how to handle them?\\
&13.3.b.9 Given the context of use, do your instructions for use contain guidelines on how to handle input data that are not comparable with the data on which the system has been tested and validated (e.g., in terms of sensitive attributes)?\\
&\cellcolor[HTML]{efefef}\textit{For completeness, please refer to Art.9 (2) and Guiding Questions Art.10.2(f-g).}\\
\midrule
\multirow{4}{0.4\textwidth}{\parbox{0.4\textwidth}{3(b)(iv). where applicable, the technical capabilities of the system to provide information to explain its output;}}
&13.3.b.10 Do the instructions for use include the technical capabilities and limitations of the AI system to explain its output (e.g., Explainable AI techniques)?\\
&13.3.b.11 Does the document contain instructions on how to intepret and comprehend the output of the system (e.g. how to interpret different visualizations given by the interpretability techniques used)?\\
&13.3.b.12 If not included, why do you think it is not applicable to your context of use?\\
&13.3.b.13 If not included, what are the foreseeable consequences on an appropriate human oversight (Art.14)?\\
\midrule
\multirow{4}{0.4\textwidth}{\parbox{0.4\textwidth}{3(b)(v). when appropriate, the performance of the system regarding specific persons or groups of persons on which the system is intended to be used;}}
&13.3.b.14 If appropriate in your application, do the instructions include the performance of your system regarding specific persons or groups of persons on which the system is intended to be used?\\
&13.3.b.15 If not appropriate, did you state it in the instructions of use?\\
&13.3.b.16 Do the instructions for use detail the means of communication in which the deployer will be informed in the case of a curriculum processing that does not belong to the group of persons for which the systems has been tested and validated?\\
&13.3.b.17 Do your instructions for use detail what to do in case of a curriculum processing that does not belong to the group of persons for which the system is tested and validated?\\
\midrule
\multirow{7}{0.4\textwidth}{\parbox{0.4\textwidth}{3(b)(vi). when appropriate, specifications for the input data, or any other relevant information in terms of the \textbf{[training]}, validation and testing data sets used, taking into account the intended purpose of the high-risk AI system;}}
&13.3.b.18 If appropriate, do the instructions cover the characteristics, capabilities, and limitations of your input data, validation, and testing data, based on the system’s intended purpose?\\
&13.3.b.19 Do the instructions for use contain information about the modalities of input data that can be processed by the AI System (e.g., text, image, video and combinations thereof)?\\
&13.3.b.20 Do the instructions for use contain specifications about different character sets of textual input data (e.g., Latin alphabet, Greek script, Cyrillic, Chinese)?\\
&13.3.b.21 Do the instructions for use contain limitations about the input data formats (e.g. document layout and length and video formats)?\\
&13.3.b.22 Do the instructions for use include limitations of the quantity and representativeness of the input data in terms of special categories of personal data?\\
&13.3.b.23 If not included, why do you think it is not appropriate for your context of use?\\
&\cellcolor[HTML]{efefef}\textit{Please refer to Art. 10.}\\
\midrule
\multirow{3}{0.4\textwidth}{\parbox{0.4\textwidth}{3(b)(vii). where applicable, the information to enable the deployers to interpret the output of the high-risk AI system and to use it appropriately.}}
&13.3.b.24 If applicable to your application, do the instructions include information for interpreting the output of the system and for using it appropriately by deployers?\\
&-  If not applicable, justify why.\\
&\cellcolor[HTML]{efefef}\textit{Please refer to Guiding Questions Art.13.1 and document them in the Instructions for Use. If not applicable, justify your motivations.}\\
\midrule
\multirow{5}{0.4\textwidth}{\parbox{0.4\textwidth}{3(c). the changes to the high-risk AI system and its performance which have been predetermined by the provider at the moment of the initial conformity assessment, if any;}}
&13.3.c.1 Are there any changes to your AI system which are predetermined by the provider at the moment of initial conformity assessment?\\
&- If yes, are these changes and the predetermined performances included in the instructions?\\
&- Do the instructions include and fully state how the provider and the deployer should interact with the system after (or in the meantime) such changes have been made?\\
&- Do the instructions for use contain information about how performance and fairness assessments will be impacted by the implementation of predetermined bias mitigation techniques not yet in place at the time of the initial conformity assessment?\\
&- Do the instructions for use contain information on performance changes when additional CV language support is added (predetermined but not yet in place at the time of the initial conformity assessment)? \\
\midrule
\multirow{6}{0.4\textwidth}{\parbox{0.4\textwidth}{3(d). human oversight measures, including technical measures, to facilitate the interpretation of the output of the AI system by deployers (Article 14);}}
&13.3.d.1 Do the instructions include human oversight measures? Describe them.\\
&13.3.d.2 Do the instructions include the technical measures to facilitate the interpretation of the output by deployers? Describe them.\\
&13.3.d.3 Do you address the problem of “interpretability” (or comprehensibility, legibility) of the output by the deployers?\\
&13.3.d.4 Do the instructions for use include information on the technical measures implemented to verify the faithfulness and robustness of interpretability techniques in supporting the interpretation of the system’s output?\\
&13.3.d.5 Do the instructions for use provide guidance on how to interpret results in cases where candidate scores are nearly identical, particularly to support the recognition of nuanced distinctions among them?\\
&\cellcolor[HTML]{efefef}\textit{For completeness, please refer to guiding questions in Art.14 and document accordingly.}\\
\midrule
\multirow{4}{0.4\textwidth}{\parbox{0.4\textwidth}{3(e). the needed computational and hardware resource specifications, the expected lifetime of the system, maintenance and care measures, including software updates, and their frequency;}}
&13.3.e.1 Do the instructions include the needed hardware specifications?\\
&13.3.e.2 Do the instructions include maintenance and care measures, including software updates, and their frequency?\\
&13.3.e.3 Do the instructions include the minimal hardware requirements that the deployer would need in order to use the AI system appropriately?\\
&13.3.e.4 Do the instructions include the expected lifetime of the AI system and what to do after the expiration date?\\
\midrule
\multirow{4}{0.4\textwidth}{\parbox{0.4\textwidth}{3(f). if relevant, the description of the mechanisms included within the high-risk AI system that allows deployers to properly collect, store and interpret the logs (Art.12).}}
&13.3.f.1 Does your high-risk AI system allow deployers to collect, store and interpret logs?\\
&- If yes, do the instructions include a description of mechanisms regarding these?\\
&- If yes, are logs written in a legible way?\\
&\cellcolor[HTML]{efefef}\textit{For completeness, please refer to Art.12.}\\
\bottomrule 

\end{longtable}
    \par\mbox{}\clearpage
\begin{table}[h]
    \centering
    \begin{tabular}{m{0.02\textwidth}m{0.98\textwidth}}
    \rowcolor[HTML]{cfe2f3}
         \multicolumn{2}{c}{
         \href{https://eur-lex.europa.eu/eli/reg/2024/1689/oj/eng\#art_14}{\large Art. 14: Human oversight}} \\ 
         & \\
         \cellcolor[HTML]{cfe2f3} \rotatebox{90}{\textbf{Interpretation guidelines}} & 
         \href{https://eur-lex.europa.eu/eli/reg/2024/1689/oj/eng#rct_73}{Recital 65}: 
         \textit{High-risk AI systems should be designed and developed in such a way that natural persons can oversee their functioning, ensure that they are used as intended and that their impacts are addressed over the system’s lifecycle. To that end, appropriate human oversight measures should be identified by the provider of the system before its placing on the market or putting into service. In particular, where appropriate, such measures should guarantee that the system is subject to in-built operational constraints that cannot be overridden by the system itself and is responsive to the human operator, and that the natural persons to whom human oversight has been assigned have the necessary competence, training and authority to carry out that role. It is also essential, as appropriate, to ensure that high-risk AI systems include mechanisms to guide and inform a natural person to whom human oversight has been assigned to make informed decisions if, when and how to intervene in order to avoid negative consequences or risks, or stop the system if it does not perform as intended. [...] 
         }
         \\ 
    \end{tabular}
\end{table}

\begin{longtable}{p{0.4\textwidth}p{0.6\textwidth}}
\rowcolor[HTML]{cfe2f3}\textbf{Compliance checklist} & \textbf{Guiding questions} \\
& \\
\endhead
\multirow{5}{0.4\textwidth}{\parbox{0.4\textwidth}{1. The high risk AI systems is designed and developed to enable effective oversight by natural persons during their operational use.}} &14.1.1 Is the high-risk AI system designed to ensure that a human remains in control of individual decision-making processes? \\
&14.1.2 Who is responsible for human oversight (e.g., developer, recruiter, or another designated role)? \\
&14.1.3 At which stages of the system pipeline does human oversight take place (e.g., after each critical step, only at the end of the process)? \\
&14.1.4 What information is provided to the human overseer to support decision-making in each individual case? \\
&14.1.5 How is the interaction between the system and the human overseer structured? Has an appropriate user interface been designed and implemented? \\
\midrule
\multirow{2}{0.4\textwidth}{\parbox{0.4\textwidth}{2. Human oversight measures are such to \textbf{prevent or minimise the risks to health, safety or fundamental rights} that may emerge when a high-risk AI system is used in accordance with its intended purpose or under conditions of reasonably foreseeable misuse}} &14.2.1 In light of the risks identified under Article 9 (including reasonably foreseeable misuse), which human oversight measures are implemented to prevent or minimise risks to health, safety, or fundamental rights? Please justify your choices by mapping each identified risk (or misuse scenario) to the corresponding oversight measure. \\
&14.2.2 Are human overseers adequately trained to recognise these risks and apply appropriate oversight measures? \\
\midrule
\multirow{5}{0.4\textwidth}{\parbox{0.4\textwidth}{3. Human oversight measures commensurate with the risks, level of autonomy and context of use of the high-risk AI system are ensured through either one or both of the following approaches: \newline 
(a) measures identified and \textbf{built by the provider} before it is placed on the market or put into service; \newline
(b) measures identified by the provider before putting it on the market and are appropriate to be \textbf{implemented by the deployer}.} }
&14.3.1 Prior to placing the system on the market or putting it into service, has the provider identified human oversight measures proportionate to the system’s risks, level of autonomy, intended purpose, and context of use? \\
&14.3.2 Have these oversight measures been embedded into the system design (e.g., through mandatory human intervention steps) before market placement? If so, describe and justify them. \\
&14.3.3 If certain oversight measures are to be implemented by the deployer rather than the provider, are they appropriate and feasible? Please document and justify your methodology. \\
&14.3.4 Has the provider supplied instructions to the deployer on how to implement these oversight measures appropriately? Please specify. \\
&14.3.5 Is there an ongoing feedback mechanism between provider and deployer to refine oversight measures as new risks emerge? \\

\midrule
\multicolumn{2}{l}{\cellcolor[HTML]{efefef} \parbox{\textwidth}{\textit{4. For the purpose of implementing paragraphs 1, 2 and 3 of Art. 14, the high-risk AI system shall be provided to the deployer in such a way that natural persons to whom human oversight is assigned are enabled, as appropriate and proportionate:}}} \\

\multirow{6}{0.4\textwidth}{\parbox{0.4\textwidth}{4(a) to \textbf{properly understand its capabilities and limitations} and duly monitor its operation, including in view of detecting and addressing anomalies, dysfunctions and unexpected performance}} 
&14.4.a.1 Have human overseers received documentation or training on the system’s functioning and intended purpose, including guidance on recognising when input data or contextual conditions may be unsuitable? \\
&14.4.a.2 What level of AI literacy is required for human overseers to participate effectively in decision-making and monitoring? \\
&14.4.a.3 Which anomalies, dysfunctions, or unexpected performance events must be reported to human overseers? \\
&14.4.a.4 How are human overseers informed of such anomalies, dysfunctions, or unexpected performance? \\
&14.4.a.5 In the event of anomalies or failed tests, what corrective interventions are implemented across the AI system pipeline (from data processing to final output)? \\
&14.4.a.6 What is the frequency of monitoring activities, and how does it ensure timely and accurate detection of anomalies or dysfunctions? \\
\midrule \multirow{6}{0.4\textwidth}{\parbox{0.4\textwidth}{4(b) to remain aware of the possible tendency of \textbf{automatically relying or over-relying on the output} produced by a high-risk AI system (automation bias), in particular for high-risk AI systems used to provide information or recommendations for decisions to be taken by natural persons}} 
&14.4.b.1 Which architectural or design choices have been implemented to mitigate automation bias (e.g. explainable AI, learning to defer, frictional AI)? \\
&14.4.b.2   Are specific human–AI interaction measures adopted to reduce over-reliance (e.g., progressive disclosure, provision of a set of alternative outputs rather than single predictions)? \\
&14.4.b.3 What information is provided to the overseer to mitigate automation bias (e.g., predictions, confidence scores, explanations)? \\
&14.4.b.4 Is the user interface designed to support effective oversight and reduce automation bias? \\
&14.4.b.5 Is oversight conducted by more than one individual per case? If so, how are divergent assessments handled? \\
&14.4.b.6  Are human overseers made aware of the risks that automation bias may pose to affected individuals, (e.g., reliance on synthetic scores without verifying robustness, illegal discrimination, or consistency of explanations)? \\
\midrule \multirow{5}{0.4\textwidth}{\parbox{0.4\textwidth}{4(c) to correctly \textbf{interpret its output} taking into account, for example, the interpretation tools and methods available;}} 
&14.4.c.1 What information and interpretability tools (e.g., feature importance, counterfactuals, examples and counterexamples) are provided to enable human overseers to interpret system outputs properly? \\
&14.4.c.2 How are human overseers trained to achieve sufficient AI literacy to interpret outputs and explanations correctly? \\
&14.4.c.3 Are explanations generated using third-party GPAI model(s)? \\
&- If so, have you considered that such explanations may be unreliable and potentially introduce systemic risk (Code of Practice, Chapter on Safety and Security, Appendix 1.4)? How is this risk mitigated? Please document and justify. \newline 
- Which technical and organisational transparency measures have been implemented to ensure that human overseers can clearly and reliably interpret the outputs of the AI system, including intermediate or final results generated by integrated third-party GPAI models?\\
&14.4.c.4 What alternative explainability (XAI) methods could be adopted? \\
\midrule \multirow{9}{0.4\textwidth}{\parbox{0.4\textwidth}{4(d) to decide, in any particular situation, \textbf{not to use} the high-risk AI system or to otherwise disregard, override or reverse the output of the high-risk AI system;}} 
&14.4.d.1 In which situations would it be appropriate not to use the high-risk AI system? \\
&14.4.d.2 Does the system include mechanisms (e.g., learning to defer) that enable it to recognise high uncertainty or risk and automatically abstain from providing any output (e.g., when the confidence score is low, the input data is not consistent with the target population, etc.)? \\
&14.4.d.3 In which situations would it be appropriate to disregard, override, or reverse the system’s output? \\
&14.4.d.4 How are human overseers alerted in such situations? \\
&14.4.d.5 How is the system designed to allow timely disregard, override, or reversal following an alert? \\
&14.4.d.6 Can human overseers override or adjust outputs beyond predefined exceptional cases (e.g., manually correcting parsing errors or adjusting rankings)? \\
&14.4.d.7 How are override or reversal actions logged and traced? \\
&14.4.d.8 Is there a feedback mechanism to integrate human evaluations into system updates when they diverge from AI outputs? \\
&14.4.d.9 Does the system automatically suspend operation when uncertainty or risk exceeds predefined thresholds? \\
\midrule 
\pagebreak\multirow{5}{0.4\textwidth}{\parbox{0.4\textwidth}{4(e) to intervene in the operation of the high-risk AI system or \textbf{interrupt the system} through a ``stop'' button or a similar procedure that allows the system to come to a halt in a safe state}} 
&14.4.e.1 How is the system interruption procedure (e.g., a “stop” button) designed? Please describe it. \\
&14.4.e.2 When system operation is halted due to malfunction, how are previously processed instances handled? \\
&14.4.e.3 Are monitoring mechanisms in place to enable human overseers to detect malfunctions and manually interrupt system execution? \\
&- If so, what information is provided to enable an informed interruption decision? \\
&- How are interruption actions and subsequent mitigation measures documented and traced? \\

\bottomrule 

\end{longtable}

    \par\mbox{}\clearpage

\begin{table}[h]
    \centering
    \begin{tabular}{m{0.02\textwidth}m{0.98\textwidth}}
    \rowcolor[HTML]{cfe2f3}
         \multicolumn{2}{c}{
         \href{https://eur-lex.europa.eu/eli/reg/2024/1689/oj/eng\#art_15}{\large Art. 15: Accuracy, Robustness and Cybersecurity}} \\ 
         & \\
         \cellcolor[HTML]{cfe2f3} \rotatebox{90}{\textbf{Interpretation guidelines}}
         & \href{https://eur-lex.europa.eu/eli/reg/2024/1689/oj/eng#rct_74}{Recital 74}: 
         \textit{High-risk AI systems should perform consistently throughout their lifecycle and meet an appropriate level of accuracy, robustness and cybersecurity, in light of their intended purpose and in accordance with the generally acknowledged state of the art. [...] The expected level of performance metrics should be declared in the accompanying instructions of use. Providers are urged to communicate that information to deployers in a clear and easily understandable way, free of misunderstandings or misleading statements. [...] In cooperation with relevant stakeholders and organisation, such as metrology and benchmarking authorities, the Commission should encourage, as appropriate, the development of benchmarks and measurement methodologies for AI systems. [...]}\newline
         
         \href{https://eur-lex.europa.eu/eli/reg/2024/1689/oj/eng#rct_75}{Recital 75}:
         \textit{Technical robustness is a key requirement for high-risk AI systems. They should be resilient in relation to harmful or otherwise undesirable behaviour that may result from limitations within the systems or the environment in which the systems operate (e.g. errors, faults, inconsistencies, unexpected situations). Therefore, technical and organisational measures should be taken to ensure robustness of high-risk AI systems, for example by designing and developing appropriate technical solutions to prevent or minimize harmful or otherwise undesirable behaviour. Those technical solution may include for instance mechanisms enabling the system to safely interrupt its operation (fail-safe plans) in the presence of certain anomalies or when operation takes place outside certain predetermined boundaries. [...]}\newline
         
         \href{https://eur-lex.europa.eu/eli/reg/2024/1689/oj/eng#rct_76}{Recital 76}:
         \textit{Cybersecurity plays a crucial role in ensuring that AI systems are resilient against attempts to alter their use, behaviour, performance or compromise their security properties by malicious third parties exploiting the system's vulnerabilities. Cyberattacks against AI systems can leverage AI specific assets, such as training data sets (e.g. data poisoning) or trained models (e.g. adversarial attacks or membership inference), or exploit vulnerabilities in the AI system's digital assets or the underlying ICT infrastructure. To ensure a level of cybersecurity appropriate to the risks, suitable measures, such as security controls, should therefore be taken by the providers of high-risk AI systems, also taking into account as appropriate the underlying ICT infrastructure.}\newline

         \href{https://eur-lex.europa.eu/eli/reg/2024/1689/oj/eng#rct_77}{Recital 77}:
         \textit{Without prejudice to the requirements related to robustness and accuracy set out in this Regulation, high-risk AI systems which fall within the scope of a regulation of the European Parliament and of the Council on horizontal cybersecurity requirements for products with digital elements, in accordance with that regulation may demonstrate compliance with the cybersecurity requirements of this Regulation by fulfilling the essential cybersecurity requirements set out in that regulation. [...]To that end, the assessment of the cybersecurity risks, associated to a product with digital elements classified as high-risk AI system according to this Regulation, carried out under a regulation of the European Parliament and of the Council on horizontal cybersecurity requirements for products with digital elements, should consider risks to the cyber resilience of an AI system as regards attempts by unauthorised third parties to alter its use, behaviour or performance, including AI specific vulnerabilities such as data poisoning or adversarial attacks, as well as, as relevant, risks to fundamental rights as required by this Regulation.}
    \end{tabular}
\end{table}
\begin{longtable}{p{0.4\textwidth}p{0.6\textwidth}}
\rowcolor[HTML]{cfe2f3}\textbf{Compliance checklist} & \textbf{Guiding questions} \\
& \\
\endhead
\multirow{17}{0.4\textwidth}{\parbox{0.4\textwidth}{1. The high-risk AI system achieves an appropriate level of accuracy, robustness and cybersecurity, and performs consistently throughout its entire lifecycle.}}
& 15.1.1  Has the system been designed to ensure an appropriate level of accuracy?\\
&-  What are the technical measures in place in order to check for the level of accuracy?\\
&- What is the threshold for the appropriate level of accuracy? How did you identify it? Document and justify.\\
&-    What are the limitations of your AI system in implementing technical measures for achieving an appropriate level of accuracy?\\
& 15.1.2  Has the system been designed to be technically robust (resilient to faults, failures, inconsistencies, and unexpected situations) as defined in Recital 75? If so, what are the protocols that enhance its robustness?\\
&-     What are the protocols and technical measures that enhance its robustness?\\
&-     To what numerical threshold does the appropriate level of robustness correspond? Justify your decision.\\
& 15.1.3     Is robustness impacted by the capability limitations of the AI system (e.g., data bottlenecks with the platform, computational cost of the operation) and the need for compliance with respect to other requirements? Document your reasoning and if applicable, measure the impact.\\
& 15.1.4     Does the system maintain cybersecurity across its full lifecycle? \\
&-     Which quantitative indicator(s) and procedures did you consider to assess the appropriate level of cybersecurity? Justify your decision.\\
&-     Which threshold(s) on the previously defined indicator(s) indicate an appropriate level of cybersecurity? Justify your decision. \\
&-     Is there an impact of the cybersecurity measures on the other requirements in Section 2? If yes, document. \\
&-     Which limitations does your AI system present with respect to cybersecurity measures and procedures? \\
& 15.1.5     Do you ensure compliance with the main cybersecurity regulation (i.e., NIS2)? How? Present and justify the normative integration. \\
& 15.1.6     Does the design of your AI system allow controlling whether an appropriate level of accuracy, robustness, and cybersecurity is sustained throughout the lifecycle? If yes, document and justify your design choices. \\
& 15.1.7     Which are the technical measures that guarantee that appropriate levels of accuracy, robustness, and cybersecurity are sustained? Justify your choice and describe the measures. \\
& 15.1.8     Which are the limitations (e.g.,  frequency) of these consistency checks? What is the impact of these limitations on sustaining an appropriate level of accuracy, robustness, and cybersecurity? \\
\midrule
\multirow{5}{0.4\textwidth}{\parbox{0.4\textwidth}{2. The benchmark proposed for testing accuracy and robustness of your high-risk AI system is aligned with the standards recognized by the Commission.}} 
& 15.2.1     Have you checked for existing EU-approved benchmarks or metrics for your AI system?\\
& 15.2.2     Are your methodology and technical measures aligned with recognized standards or practices? \\
& 15.2.3     Does your design allow for the update of the metrics if new EU benchmarks are issued? \\
& 15.2.4     If you adopted benchmarks different from the recognized standards (for limitations or any specific disagreement), did you justify how your choice still remains aligned with official standards? \\
& 15.2.5  If there are no recognized benchmarks, how do you justify your own benchmark adoption (comparing with state-of-the-art in the academic literature etc.)? \\
\midrule
\multirow{5}{0.4\textwidth}{\parbox{0.4\textwidth}{3. The system’s accuracy and the relevant accuracy metrics used to assess it are declared in the instructions for use.}}
& 15.3.1  Have you specified the accuracy levels of the system and documented these in the instructions for use? \\
& 15.3.2  Are the accuracy metrics documented in the instructions for use? \\
& 15.3.3  Is the documentation regarding accuracy levels and accuracy metrics understandable by the deployer? \\
&- Is the documentation regarding accuracy levels and accuracy metrics understandable by the human overseer (i.e., recruiter)? \\
& 15.3.4  If the AI system has been developed by integrating third-party GPAI models, what transparency measures ensure the model's performance metrics are disclosed, and what technical measures and contractual safeguards guarantee that the accuracy levels remain appropriate for the system's intended purpose? \\
& \cellcolor[HTML]{efefef}\textit{Please refer to Art.13.3(b)(ii)).}\\
\midrule
\multirow{16}{0.4\textwidth}{\parbox{0.4\textwidth}{4. (Part I) Technical and organizational measures are taken to guarantee that the system is as resilient as possible regarding errors, faults, or inconsistencies, whether these originate \textbf{within the system itself} or from the \textbf{environment} in which it operates, in particular due to its interaction with \textbf{people} and \textbf{other systems}.}}
& 15.4.1  What classes of failure (e.g., software crashes, network outages, hardware faults) has the system been tested against, and how are they detected at runtime?\\
& 15.4.2   How do you define the system’s acceptable mean-time-to-failure and mean-time-to-repair targets and who is accountable for them?\\
& 15.4.3    Which are the events that could trigger within-system and/or external errors, faults, or inconsistencies in your AI system?\\
& 15.4.4   Which are the organizational and technical measures that make the system resilient to external errors, faults, or inconsistencies (e.g., unexpected user inputs, unexpected change of context)? Describe them and justify your choice.\\
&-     Has the system been tested against external errors, faults, or inconsistencies?\\
&- How are external errors, faults, or inconsistencies detected at runtime?\\
&- For each possible external system error, fault, or inconsistency, what is the course of action that you designed after detection?\\
& 15.4.5  Which are the organizational and technical measures that make the system resilient to within-system errors, faults, and inconsistencies (e.g., rapid repeated queries, unusual use patterns by authorized persons)? Describe them and justify your choice.\\
&- Has the system been tested against within-system errors, faults, and inconsistencies?\\
&- How are within-system errors, faults, and inconsistencies swiftly detected at runtime?\\
&- For each possible within-system error, fault, or inconsistency, what is the course of action that you designed after detection?\\
& 15.4.6     Did you identify any event that could trigger errors, faults, or inconsistencies due to the interaction of your AI system with other systems? Describe them.\\
&- Has the system been tested against the identified events?\\
&- How are these events detected at runtime?\\
&- For each possible event, what is the course of action that you designed?\\
& 15.4.7     Did you identify any event that could trigger errors, faults, or inconsistencies due to errors by third-party systems on which your system depends on?\\
&- Has the system been tested against the identified events?\\
&- How are these events detected at runtime?\\
&- For each possible event, what is the course of action that you designed?\\
\midrule
\multirow{7}{0.4\textwidth}{\parbox{0.4\textwidth}{4. (Part II) Robustness is achieved through technical redundancy solutions, which may include backup or fail-safe plans.
[Not Legally Binding]}}
& 15.4.8  Have technical redundancy solutions (e.g., fail-safes, backups) been implemented to ensure robust and uninterrupted operation if one component fails? \\
&-  Are redundancy solutions documented for each component of the pipeline? Justify your choice of redundancy solutions and argue why they guarantee the appropriate level of robustness.\\
& 15.4.9  Where do you maintain redundant copies of critical models, data, and services?\\
&-  Who is responsible for the appropriate maintenance of these redundant copies?\\
& 15.4.10  How often are backups performed, and how do you verify backup integrity and restore procedures?\\
&-  Who is responsible for the backup operation and for the verification of backup integrity and restore procedures?\\
& 15.4.11  What capacity-planning forecasts ensure that redundant resources can handle peak load if the primary path fails? Document and justify your expectations.\\
& 15.4.12  If the AI system has been developed by integrating third-party GPAI models, what transparency measures regarding the integrated model's limitations, what technical measures for redundancy (such as fail-safe plans), and what contractual safeguards ensure that the robustness of the system is maintained against faults or unexpected situations? \\
\midrule
\multirow{13}{0.4\textwidth}{\parbox{0.4\textwidth}{4. (Part III) If the AI system continues to learn after deployment, it shall eliminate or reduce as far as possible the risk of possibly biased outputs influencing inputs for future operations \textbf{(feedback loops)}, and ensure that any such feedback loops are duly addressed with appropriate mitigation measures.}}
& 15.4.13  Is your system learning after deployment? \\
& 15.4.14  Which inputs are used for learning after deployment (e.g., machine outputs, human feedback, monitoring and decisions)? \\
& 15.4.15  How do you assess the effect that such inputs may have on the quality of data eventually leading to a change in fairness assessments?\\
& 15.4.16  How are input data used by the AI system (i.e., in addition to processing, for instance for post-deployment learning)?\\
& 15.4.17  Can you describe the learning procedure where outputs are used as inputs in future operations? \\
&-  With which timing is this operation performed? \\
&- How do you ensure that induced bias is detected swiftly so as to ensure an appropriate level of robustness? \\
& 15.4.18  If the system continues to learn post-deployment, how do you ensure that it prevents or reduces low-quality outputs from feeding future learning?\\
&-  Does your design include mitigation strategies? If yes, justify your choice.\\
&-  How do you assess if a mitigation strategy has been successful or not?\\
&-  What is your designed course of action in case the mitigation strategy fails?\\
& 15.4.19     What are the technical and organizational safeguards put in place so as to guarantee that the new input data is not poisoned (e.g., due to tampering by authorized or unauthorized personnel)?\\
& 15.4.20    What are the technical and organizational safeguards put in place so as to guarantee that the model has not been tampered with during the feedback loop procedure (e.g., due to adversarial attacks)?\\
\midrule

\multirow{24}{0.4\textwidth}{\parbox{0.4\textwidth}{5. The system is protected against unauthorized third parties to alter its use, outputs, or performance by exploiting system vulnerabilities. The technical solutions aiming to ensure the cybersecurity shall be appropriate to the relevant circumstances and risks. Technical solutions to address AI-specific vulnerabilities shall include, where appropriate, measures to prevent, detect, respond to, resolve, and control for attacks trying to manipulate the training data set (data poisoning), pre-trained components used in training (model poisoning), input designed to cause the AI model to make mistakes (adversarial examples or model evasion), confidentiality attacks, or model flaws.}}
& 15.5.1  Which internal and external threat actors and attack vectors have you identified?\\
& 15.5.2  How have you classified cybersecurity risks by severity and likelihood, and how often are the risks and their severity and likelihood in your threat model re-evaluated?\\
& 15.5.3  What are the technical measures (e.g., anomaly detection on incoming data streams) that you use to spot unusual patterns indicative of poisoning attempts? \\
& 15.5.4  If you incorporate third-party or pre-trained components, how do you check the integrity of such components? \\
& 15.5.5    Have you implemented protocols for security and deception risks arising from the input data at inference time, e.g., adversarial examples and white-fonting attacks? \\
& 15.5.6   How are updates to pre-trained models authenticated, versioned, and logged to prevent silent insertion of malicious weights? \\
& 15.5.7  Which adversarial-robustness techniques (i.e., adversarial training, input preprocessing) are integrated to protect the model against adversarial attacks? \\
& 15.5.8  Do you run automated penetration tests targeting AI-specific vulnerabilities? \\
&-  How often do you run them? Justify your decision. \\
&-  Are the test procedures and logs documented? \\
&-  Who is responsible for the penetration tests and their appropriate documentation? \\
& 15.5.9  In case of an incident (e.g., data poisoning, model poisoning), are logs and data securely collected to allow an appropriate post-incident analysis? \\
& 15.5.10  Are access controls in place to prevent unauthorized use or modification of the system? \\
& 15.5.11  Are monitoring systems in place to detect tampering or misuse? \\
& 15.5.12  Have you identified and addressed model flaws such as inherent design vulnerabilities? \\
& 15.5.13  Do you have technical and organizational measures put in place to appropriately address confidentiality attacks (e.g., leakage of training data)? Describe them. \\
& 15.5.14  Is a cybersecurity strategy in place, adapted to the context and risk profile of the system? Describe it and justify your design. \\
& 15.5.15  Which cybersecurity measures are in place to prevent cybersecurity attacks? Describe them and justify their appropriateness. \\
& 15.5.16  Which cybersecurity measures are in place to detect incidents? Describe them and justify their appropriateness. \\
& 15.5.17 Which cybersecurity measures are in place to respond to breaches? Describe them and justify their appropriateness. \\
& 15.5.18   Which cybersecurity measures are in place to resolve threats? Describe them and justify their appropriateness. \\
& 15.5.19  Which cybersecurity measures are in place to control and monitor the system continuously? \\
& 15.5.20  Are cybersecurity risks periodically reassessed (e.g., after updates or retraining)? If yes, how? \\
& {15.5.21  If the AI system has been developed by integrating third-party GPAI models, what transparency measures regarding model vulnerabilities, what technical measures to prevent or detect AI-specific attacks (like data poisoning), and what contractual safeguards are in place to ensure cybersecurity throughout the lifecycle of the AI system?} \\
& \cellcolor[HTML]{efefef}\textit{Regarding cybersecurity aspects, for completeness refer to cybersecurity law (i.e., NIS2).} \\
\bottomrule 

\end{longtable}


\end{document}